\documentclass[12pt,preprint]{aastex}
\shorttitle{}
\shortauthors{Nesvorn\'y}
\begin{document}
\title{Dark Primitive Asteroids Account for a Large Share \\ of K/Pg-Scale Impacts on the Earth}
\author{David Nesvorn\'y, William F. Bottke, Simone Marchi}
\affil{Department of Space Studies, Southwest Research Institute, 1050 Walnut St., \\Suite 300, 
Boulder, CO 80302, USA} 
\begin{abstract}
A dynamical model for large near-Earth asteroids (NEAs) is developed here to understand the occurrence 
rate and nature of Cretaceous-Paleogene (K/Pg) scale impacts on the Earth. We find that 16--32 (2--4) 
impacts of diameter $D>5$ km ($D>10$ km) NEAs are expected on the Earth in 1 Gyr, with about a half of 
impactors being dark primitive asteroids (most of which start with semimajor axis $a>2.5$ au). These 
results explain why the Chicxulub crater, the third largest impact structure found on the Earth (diameter 
$\simeq180$ km), was produced by impact of a carbonaceous chondrite. They suggest, when combined with
previously published results for small ($D \lesssim 1$ km) NEAs, a size-dependent sampling of the 
main belt. We conclude that the impactor that triggered the K/Pg mass extinction $\simeq 66$ Myr ago 
was a main belt asteroid that quite likely ($\simeq 60$\% probability) originated beyond 2.5 au. 

\end{abstract}
\section{Introduction}
The Earth impact record accounts for $\sim$200 recognized crater structures and $\sim$50 deposits (Schmieder 
and Kring, 2020). This collection is largely incomplete and contains severe biases. The impact age distribution, 
inferred from isotopic and stratigraphic analyses, shows that a great majority of preserved terrestrial craters 
formed in the last $\sim$650 Myr (Mazrouei et al., 2019; Keller et al., 2019). 
Old impact structures on non-cratonic terrains were apparently erased by tectonic recycling of the 
crust, erosion, and buried under layers of sediments and lava (e.g., Grieve 1987). The Archean spherule beds, 
with two impact clusters at $\sim2.5$ Ga and $\sim3.4$ Ga, are a reminder that the early bombardment must 
have been intense (Johnson and Melosh, 2012; Bottke et al., 2012; Johnson et al., 2016; Marchi et al. 2021). 

Isotopic compositions and elemental abundances in impact melt rock and glass samples can be used to 
determine the nature of impactors. The overall record is dominated by impactors with a composition similar 
to ordinary chondrites (OCs; e.g., Koeberl et al., 2007). This is consistent with theoretical models of 
near-Earth objects and terrestrial impactors (Bottke et al. 2002, Granvik et al. 2018), which show that 
a great majority of terrestrial impactors start in the innermost part of the asteroid belt where asteroids 
often have the OC composition (as inferred from spectrophotometric observations; DeMeo et al., 2015). 

Interestingly, the famous Chicxulub crater (Hildebrand et al., 1991), that has been linked with the 
Cretaceous-Paleogene (K/Pg) mass extinction (Alvarez et al., 1980; Chiarenza et al., 2020), does not 
follow the suit. Here, chromium found in sediment samples taken from different K/Pg boundary sites (Shukolyukov 
and Lugmair, 1998; Trinquier et al., 2006), as well as a meteorite found in K/Pg boundary sediments from 
the North Pacific Ocean (Kyte et al., 1998), suggest the impactor was a carbonaceous chondrite (CC). This 
is quite rare: among dozens of terrestrial craters with inferred impactor composition (Schmieder and Kring, 
2020), a CC impactor was suggested only for Lonar and Zhamanshin (e.g., Mougel et al., 2019). The nature 
of the Chicxulub crater impactor -- the third largest preserved impact crater on the Earth (diameter 
$D \simeq 180$ km; after Sudbury and Vredefort) -- may thus suggest a special circumstance (e.g., Bottke 
et al., 2007).

Here we study the origin of large terrestrial impactors to determine where, in the main belt, they come 
from, and how often they have the CC composition. For that, we construct a dynamical model of large near-Earth 
asteroids (NEAs) based on new simulations that follow diameter $D>5$ km asteroids from their original 
orbits in the main belt. The WISE albedos ($p_{\rm V}$) are taken as a proxy for the CC composition ($p_{\rm V} \leq 0.1$) 
of main belt asteroids (MBAs) and NEAs (Mainzer et al., 2011a). The number of impacts on the terrestrial worlds
is directly recorded by the $N$-body integrator (approximate \"Opik schemes are not used here). We find
that CC asteroids should be responsible for $\simeq 50$\% of large impacts on the Earth, and discuss the 
implications of our work for the occurrence rate and nature of K/Pg-scale impactors. 
  
\section{Many-Source NEO Models} 

The near-Earth objects (NEOs) have a relatively short life cycle ($\sim0.1$-100 Myr; Gladman et al., 1997). 
They impact planets, disintegrate near the Sun, and end up being ejected from the Solar System (Farinella et 
al., 1994; Granvik et al., 2016). To persist over billions of years, the NEO population must be repopulated 
from MBAs and comets (Bottke et al. 2002). It is estimated that there are roughly 1,000 NEOs with diameters 
$D>1$ km (e.g., Harris \& D'Abramo, 2015; Morbidelli et al., 2020). Thus, for the NEO 
population to be in a steady state, a new $D>1$ km asteroid/comet must evolve onto a NEO orbit once every 
$\simeq4,000$ yr (average NEO lifetime $\simeq 4$ Myr assumed here; Bottke et al., 2002). Several dynamical 
models were developed to investigate this process and determine the contribution of different sources to NEOs 
(e.g., Bottke et al. 2002; Morbidelli et al. 2002, 2020; Greenstreet et al., 2012; Granvik et al., 2018). 

These models have a common ground. A large number of test bodies is placed onto source orbits and an 
$N$-body integrator is used to follow them as they become NEOs. Bottke et al. (2002; hereafter B02) 
adopted five NEO sources: the $\nu_6$ resonance, intermediate Mars crossers (IMCs), 3:1 resonance, outer main 
belt and Jupiter Family Comets (JFCs), whereas Granvik et al. (2018; hereafter G18) -- striving for 
improved model accuracy -- used 7 to 23 different source regions. A {\it target} NEO region is defined, 
where the model distribution of NEOs is analyzed in detail (perihelion distance $q<1.3$ au and 
semimajor axis $a<4.2$~au in both B02 and G18). The relative contribution of different sources to the 
target region is determined by accounting for observational biases and comparing the results with observations 
(e.g., the Catalina Sky Survey; Christensen et al., 2012; Jedicke et al., 2016). For example, 
B02 found that the five sources mentioned above contribute by 37\%, 27\%, 20\%, 10\% and 6\% to the NEO
population, respectively. 

Calibrated many-source NEO models have been used to estimate the observational incompleteness as a function 
of NEO orbit and size. When the albedo information is folded in, the models also provide intrinsic orbital 
distributions of dark and bright NEOs and their relative importance for impacts. For example, G18 
estimated that $\sim$80\% of terrestrial impactors with absolute magnitude $17<H<22$ ($0.17<D<1.7$ km for 
$p_{\rm V}=0.1$ or $0.12<D<1.2$ km for $p_{\rm V}=0.2$) come from the $\nu_6$ resonance at the inner edge of the main belt, where MBAs are typically 
bright ($p_{\rm V}>0.1$) and spectroscopically consistent with OCs (Binzel et al., 2004, 2019; Mainzer et al., 
2011a; Vokrouhlick\'y et al., 2017). This is presumably reflected in the terrestrial impact record (Sect. 
1). The inner belt ($2<a<2.5$ au) was found to be the main source of dark primitive impactors in 
G18.\footnote{The forward-modeling method of B02 and G18 has several 
advantages over direct attempts to remove biases from NEO observations (e.g., Stuart 2001). Notably, 
the bias removal does not work in the regions of orbital space where no or a very few objects are detected. 
Given that the number of objects discovered in any particular NEO survey is usually only a very small 
fraction of the whole population, the direct debiasing method can struggle to produce a full coverage of 
the orbital domain and describe potential correlations between different parameters.} 

The dynamical models of B02 and G18 have an important limitation: even though the source populations 
are chosen with care, this does not guarantee that some NEO sources might be missing. G18 adopted 
more sources than B02, but this is not ideal as well, because as the number of sources increases, 
the number of free model parameters increases as well, and degeneracies between neighbor sources become 
important. For example, with $n$ source populations, there are $n$ parameters that weight the individual contribution 
of sources ($n=7$-23 in G18), and $n$ additional parameters for the ratio of dark and bright objects ($n=6$ 
in Morbidelli et al., 2020). Extra parameters arise in the many-source models because 
one has to set the initial orbital distribution of bodies in each source. To help with this choice, 
Granvik et al. (2017) followed the orbits of MBAs -- as they drift by the Yarkovsky effect 
(Vokrouhlick\'y et al. 2015; the Yarkovsky effect is a radiative recoil force produced by thermal photons
emitted from asteroid surface) and enter resonances -- and used the results to inform the starting orbits 
in the NEO model (G18). 

\section{Single-Source NEA Model}

Here we develop a new, physically-grounded NEA model with fewer parameters (we ignore comets in this 
work). The model is explained as follows. Ideally, one would like to establish the relation between MBAs and NEAs 
without restricting the link to a large number of intermediate sources. Therefore, there is only one 
source in our model: the whole main belt. This removes the need for the empirical weight factors in 
the many-source models. In an ideal world, where the MBAs were characterized well enough (e.g., a 
complete sample down to some small size, known albedo/taxonomic distributions), and where we had a 
detailed and accurate understanding of the radiation effects (which feed MBAs into escape 
resonances; e.g., Morbidelli and Vokrouhlick\'y, 2003), a fully {\it physical} NEA model could be 
developed. In a physical model, all model parameters would be related to physical quantities such as 
the distribution of MBA spins and shapes, their thermal properties, etc.

In practice, however, a number of issues can compromise such an ambitious effort. For example, we 
do not have a complete understanding of how the MBA spin vectors are affected by the YORP effect 
(Rubincam, 2000; Bottke et al. 2006a, Vokrouhlick\'y et al., 2015; the YORP effect is a radiative 
recoil torque that affects asteroid rotation), collisions (Holsapple, 2021) and 
spin-orbit resonances (Vokrouhlick\'y et al., 2003). The YORP effect for an individual body depends 
on its overall shape, and is sensitive to small shape changes (Statler, 2009; e.g., generated by 
impacts or landslides). To realistically model the YORP effect for a statistically large number of 
MBAs, where no such detailed information is available, it is therefore preferable to parametrize 
the YORP strength relative to a standard (\v{C}apek \& Vokrouhlick\'y, 2004; Vokrouhlick\'y et al., 
2006; Lowry et al., 2020). The effect of the lateral heat conduction must be accounted for (Golubov 
\& Krugly, 2012). Additional YORP-related complications arise because we do not know what happens 
when asteroids reach very fast or very slow rotation rates (i.e., fast rotation may lead to mass shedding, 
while very slow rotation rates may lead to tumbling), and how this feeds back 
to changing the YORP torques. 

All these considerations have a major importance for the orbital evolution of MBAs. This is because the 
Yarkovsky drift rate depends on asteroid's obliquity $\theta$ (the diurnal Yarkovsky force is proportional 
to $\sin \theta$), and as the obliquity changes due to YORP and collisions, the Yarkovsky drift rate 
changes as well. Thus, to realistically replicate how MBAs reach resonances and escape from the main 
belt, these complex relationships would need to be taken into account. This is especially critical for 
small MBAs that must undergo many YORP cycles  -- i.e., reach very slow or very fast rotation -- before 
they could evolve from the main belt. The small MBAs also have short collisional lifetimes (Bottke et 
al. 2005) and can be disrupted before they could reach NEO orbits. Here we therefore only consider 
{\it large} MBAs ($D>5$ km), for which many of the complicating factors discussed above can be brushed 
aside.

\section{Methods}

\subsection{Large MBA Selection}
 
We used the Wide-field Infrared Survey Explorer (WISE) catalog (Mainzer et al., 2019) to select all known 
$D>5$ km MBAs, 42721 in total, with $a<4.5$ au and $q=a(1-e)>1.7$ au. This is {\it not} a complete sample. 
In fact, the WISE catalog of MBAs is practically complete only for $D \gtrsim 7$ km (Mainzer et al. 2011b). 
The problem of observational incompleteness can mainly be important for the outer belt. The WISE detections, 
however, are relatively insensitive to the visible albedo (Mainzer et al. 2015) and the selected sample 
is therefore {\it not} (strongly) biased toward high albedo asteroids (see below). 

The diameter cutoff that 
we use here is a compromise between several different factors. We want to investigate large asteroids to 
limit the influence of various uncertainties discussed in Sect. 3, and because our main goal is to understand 
the origin of the K/Pg impactor. We do not want, however, to limit the scope of this study only to 
$D \gtrsim 10$ km, the suggested size of the K/Pg impactor from classical irridium abundance studies 
(Alvarez et al., 1980) and Chicxulub--impact modeling (e.g., Collins et al. 2020). 
There are only two known 
NEAs with $D \gtrsim 10$ km (Eros and Ganymed), and we would therefore not have any independent means to 
validate our model if only $D \gtrsim 10$ km were considered (see Nesvorn\'y and Roig, 2018).
     
Given the bimodal albedo distribution of MBAs, we define dark, low-albedo MBAs as $p_{\rm V} \leq 0.1$ 
and bright, high-albedo MBAs as $p_{\rm V} > 0.1$. There is a good correspondence between albedo and taxonomic 
type with the dark (bright) asteroids typically belonging to the C-complex (S-complex) groups (Mainzer et 
al., 2011a; Pravec et al., 2012). Here we therefore use albedo as a proxy for the nature of each body: 
$p_{\rm V} \leq 0.1$ for C-complex or CCs, and $p_{\rm V} > 0.1$ for S-complex or OCs. It is acknowledged 
that many exceptions exist to this idealized, one-to-one correspondence, and this could 
potentially affect some of the results presented in this work.

We checked that the ratio of the dark/bright MBAs, ${\rm d}/{\rm b}$, is practically the same for different 
diameter cutoffs. For example, in the whole main belt, ${\rm d}/{\rm b}=2.7$ for $D>5$ km, 2.9 for $D>7$ 
km and 2.8 for $D>10$ km. In the outer belt ($a>2.82$ au), ${\rm d}/{\rm b}=3.7$ for $D>5$ km, 3.9 for 
$D>7$ km and 3.5 for $D>10$ km. This indicates that the distribution for $D>5$ km is not strongly biased. 
If the selected sample is missing some small fraction ($\sim10$\%?) of outer dark MBAs, our work would somewhat 
underestimate the contribution of the outer dark MBAs to the NEO population. Establishing this factor is 
left for future work.    

\subsection{Yarkovsky Clones}

Three clones are considered for each selected asteroid. The first clone is given the maximum negative Yarkovsky 
drift rate, the second one is given the maximum positive Yarkovsky drift rate, and the third one is given no 
drift. The maximum negative/positive rate is assigned to each individual body depending on its 
size, albedo, and semimajor axis (Vokrouhlick\'y et al. 2015). It is informed from the measurement of the 
Yarkovsky effect on asteroid (101955) Bennu (diameter $D_{\rm Bennu} \simeq 0.5$ km, semimajor axis $a_{\rm Bennu}
=1.126$ au, obliquity $\theta \simeq 178^\circ$, taxonomic type B -- part of the C-complex): ${\rm d}a/{\rm d}t = 
(-19.0 \pm 0.1) \times 10^{-4}$ au Myr$^{-1}$ (Chesley et al. 2014). Thus, for example, for a dark C-complex MBA, 
we use clones with the maximum drift rates $|{\rm d}a/{\rm d}t| = 1.9 \times 10^{-3} (D_{\rm Bennu}/D) (a_{\rm Bennu}/a)^2$ au Myr$^{-1}$.
This is consistent with Bottke et al. (2006a) who quoted $2 \times 10^{-4}$ au Myr$^{-1}$ for a $D=1$ km MBA 
with $a=2.5$ au. When setting the drift rates for S-complex MBAs, we account for their larger densities 
and higher albedos. The {\tt swift\_rmvs4} code (Levison \& Duncan 1994) was modified to account for 
${\rm d}a/{\rm d}t$ from the Yarkovsky effect.

The Yarkovsky drift rates of individual clones were assumed to be unchanging with time. This is an important 
approximation. In reality, the YORP effect and small impacts\footnote{The collisional evolution of MBAs is 
ignored here, because large MBAs have relatively long collisional lifetimes (e.g., Bottke et al., 2005).} 
would change the spin axis vector and induce time-dependent drift rates. The YORP effect cannot be neglected 
for a $D \sim 5$ km MBA (Vokrouhlick\'y et al. 2003). It acts to evolve the obliquity 
toward $\theta=0^\circ$ and 180$^\circ$ and maximize the magnitude of the Yarkovsky drift (maximum positive for 
$\theta=0^\circ$ and maximum negative for $\theta=180^\circ$). This is the reason, in the first place, why we use 
two clones with the maximum positive and maximum negative drift rates: this should cover the full range of 
possibilities. Cases with intermediate drift rates are expected to lead to intermediate results (this is not 
demonstrated here; to demonstrate it we would need to include the intermediate drift rates in the simulation).
With three clones for each selected $D>5$ km MBA we have nearly 130,000 bodies in total. 

\subsection{Orbital Integration of $D>5$ km Asteroids}

Our numerical integrations included planets, which were treated as massive bodies that gravitationally interact 
among themselves and affect the orbits of all other bodies, and asteroid clones, which were massless (i.e., they 
did not affect each other and the planets). The integrations were performed with the {\it Swift} $N$-body 
integrator (Levison and Duncan, 1994), which is an efficient implementation of the symplectic Wisdom-Holman map 
(Wisdom and Holman, 1991). Specifically, we used the code known as {\tt swift\_rmvs4} that we adapted for the 
problem at hand. It was modified such that it can be efficiently parallelized on a large number of CPUs. The 
treatment of close encounters between planets and asteroid clones in {\tt swift\_rmvs4} is such that the evolution of 
planetary orbits on each CPU is strictly the same (and reproducible). The Yarkovsky force was included in the 
kick part of the integrator. 

The integrations were performed on the NASA's Pleiades Supercomputer. They were split over 8560 Pleiades cores 
with each core dealing with 15 clones. All planets except Mercury were included. Leaving out Mercury allowed us 
to perform the simulations for a reasonably low CPU cost. The gravitational effects of Mercury were found 
insignificant in previous studies (e.g., Granvik et al., 2016). We used a 10-day timestep and verified that the 
main results do not change when a 3-day timestep is used (Nesvorn\'y and Roig (2018) tested a 1-day timestep 
as well). The main integrations covered 1 Gyr allowing us to monitor impacts on the terrestrial planets 
during that time. They were run forward in time from the current epoch such that the results should 
be strictly applicable to the impact flux during the next 1 Gyr. Still, with some uncertainty, the impact flux 
obtained from our integration can be thought as being representative for a long-term average near the 
current epoch. 

\subsection{Model for NEAs and Impacts}

The orbits of MBAs that evolved to $q<1.3$ au were saved at fixed time intervals ($\Delta t = 1,000$ 
yr) and used to build the model orbital distribution for $D>5$ km NEAs (see B02 and G18 for details). 
The model distribution is compared with the observed orbits of $D>5$ km NEAs (Table 1) in Sect. 5.4. 
We monitored the escaping bodies in a selected time interval $\Delta T$ and recorded the total time 
spent by all model bodies on orbits with $q<1.3$ au, $T_{q<1.3}$, 
in $\Delta T$. To estimate the number of $D>5$ km NEAs in a \textit{steady state}, $n_5$, we computed 
$n_5=T_{q<1.3}/\Delta T$. We also extracted from the simulations the number of all individual bodies, 
$N_{q<1.3}$, that reached $q<1.3$~au during $\Delta T$. The {\it mean} dynamical lifetime of $D>5$ km 
asteroids in NEA space was computed as $t_5=T_{q<1.3}/N_{q<1.3}$. 

The impacts of model NEAs on Venus, Earth and Mars were recorded by the {\tt swift\_rmvs4} integrator. 
The results based on the {\it Swift}-recorded impacts are more reliable than the ones obtained from the 
\"Opik code (e.g., B02; G18). For example, the \"Opik code does not account for the resonance protection 
mechanism, which may be especially important for Mars crossers that just evolved onto NEO orbits from various 
weak mean motion resonances with Mars (Migliorini et al., 1998; Morbidelli and Nesvorn\'y, 1999).

\section{Results}

\subsection{Dynamical Loss of Large MBAs}

\begin{figure}
\epsscale{0.49}
\plotone{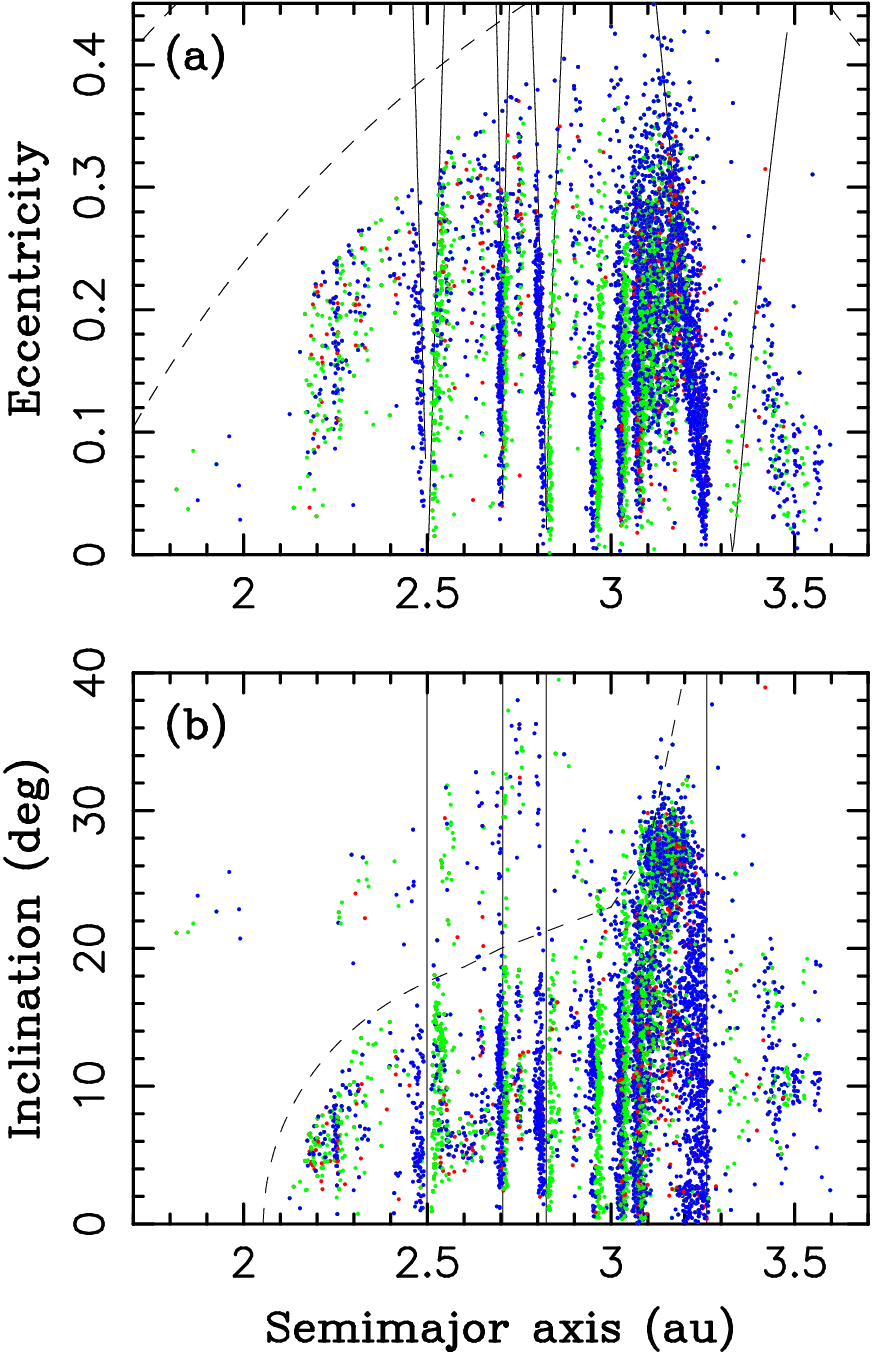}
\plotone{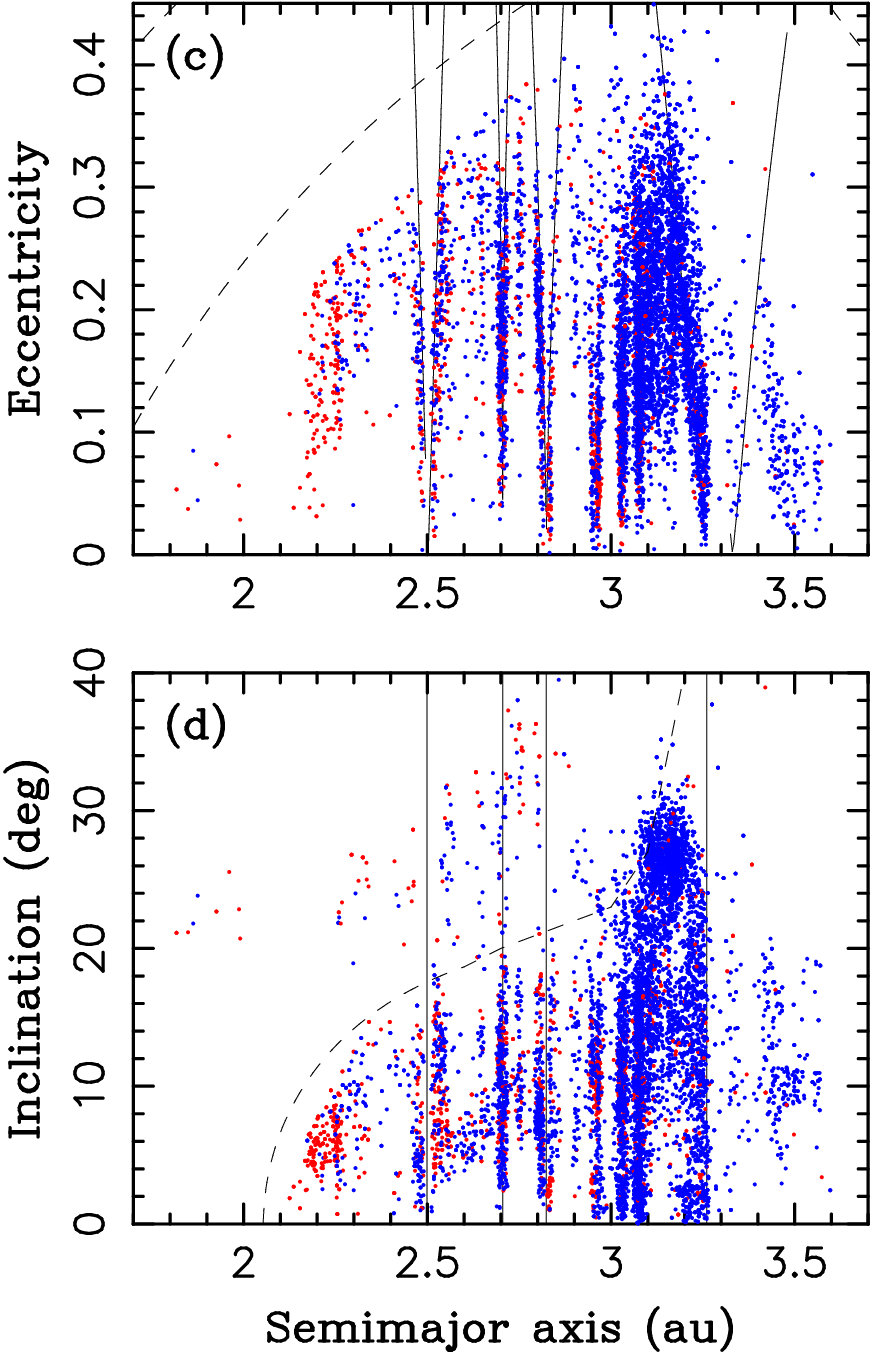}
\caption{Panels (a) and (b): The initial orbits of $D>5$ km asteroids that escaped from the main belt in 0.4 Gyr.
We only show the first 0.4 Gyr of our simulations, because the number of escaping MBAs in the model decays over 
time (Sect. 5.1), and we believe that the first 0.4 Gyr is the best choice to represent the long term average. 
The color indicates the Yarkovsky drift: ${\rm d}a/{\rm d}t=0$ (red), ${\rm d}a/{\rm d}t<0$ (green),
and  ${\rm d}a/{\rm d}t>0$ (blue). There is a clear preference for drift toward resonances. The clump of bodies
at $a=3.1$--3.2 au and $i=25^\circ$--30$^{\circ}$ is the Euphrosyne family. Panels (c) and (d): The same as (a) and 
(b) but now the color denotes the WISE albedo: blue for $p_{\rm V} \leq 0.1$ and red for $p_{\rm V} > 0.1$. The thin 
solid lines show the location of orbital resonances with Jupiter. The dashed line in (a) and (c) is the Mars 
crossing limit. The dashed line in (b) and (d) is the secular resonance $\nu_6$.}
\label{escape}
\end{figure}

The initial orbits of asteroids that escaped from the main belt in the course of our integration are highlighted
in Fig. \ref{escape}. Many more MBAs escape from the outer belt than from the inner belt, primarily because 
the outer belt represents a much larger source reservoir. For example, for $D>5$ km, there are 3373 bodies 
in the inner belt ($a<2.5$ au; 8\% of the total number of $D>5$ km MBAs), 10178 bodies in the middle belt 
($2.5<a<2.82$ au; between the 3:1 and 5:2 resonances with Jupiter; 24\%), and 29170 bodies in the outer belt 
($a>2.82$ au; 68\%). The radial distribution of $D>10$ km bodies is similar (6\%-21\%-73\%), but the overall 
number is $\simeq$5.2 times lower.\footnote{As for the distribution of dark and bright MBAs, the dark/bright ratio for 
$D>5$ km is ${\rm d/b}=1.0$, 1.7, 3.7 in the inner, middle and outer zones, respectively (${\rm d/b}=1.3$, 
1.8, and 3.5 for $D>10$ km). This adds to ${\rm d/b}=2.7$ for $D>5$ km (or ${\rm d/b}=2.8$ for $D>10$ km) in 
the whole main belt.} 
      
We now turn our attention to escape statistics. The number of MBAs escaping per 100 Myr slowly
declines over time and the decline rate depends on the assumed Yarkovsky drift. Overall, there were 3211
clones escaping in $t<100$ Myr, 2295 in $100<t<200$ Myr, 2002 in $200<t<300$ Myr, 1678 in $300<t<400$ Myr,
and 1565 in $400<t<500$ Myr (and the decline continues beyond 500 Myr). The decline is stronger when only clones 
with ${\rm d}a/{\rm d}t=0$ are considered, presumably because the escape resonances are not re-filled by new 
MBAs in this case. The decline is weaker for MBAs with the Yarkovsky drift: 1057 escape in $t<100$ Myr and 607 in 
$400<t<500$ Myr for ${\rm d}a/{\rm d}t<0$; 1140 for in $t<100$ Myr and 687 in $400<t<500$ Myr for ${\rm d}a/{\rm d}t>0$. 
Therefore, even in this case, the resonances are not refilled efficiently enough to keep the escape rate 
constant. This may either be real (i.e., the actual escape rate will decline in the future) or related to some 
dynamical effects that were not included in our model (e.g., caused by the Yarkovsky drift rate variability).
  
The escape rate of MBAs in the first 100 Myr of our simulations is considered the most realistic proxy 
for the present epoch. We find that one $D>5$ km asteroid escapes from the main belt every $\simeq$93 kyr,
which means that $\simeq$2.5\% of the whole population of $D>5$ km MBAs escape from the main belt in 100 Myr.
Extrapolated to longer timescales this would represent one $e$-fold reduction of the original 
population every $\sim$4 Gyr, which is roughly consistent with the previous estimates (Minton and Malhotra,
2010; Nesvorn\'y et al., 2017). Note, however, that this is somewhat coincidental because the previous works 
did not account for the Yarkovsky drift and adopted a random distribution of MBA orbits (including 
unstable orbits in resonances). For $D>10$ km, we determine that $\simeq1.7$\% MBAs escape in 100 Myr, 
which is identical to the estimate given in Nesvorn\'y \& Roig (2018).  

We find that 123 $D>5$ km MBA clones were eliminated from the inner belt, 540 from the middle belt, and 2548 from 
the outer belt (all for $t<100$ Myr). These numbers correspond to 1.2\%, 1.8\%, and 2.9\% of the populations 
in each zone. This means that the outer belt is more leaky than the inner belt probably because the 
outer MBAs have orbits opportunistically close to resonances at the current 
epoch.\footnote{The fractions escaping in 1 Gyr are 17.2\%, 16.6\%, and 19.0\%, respectively, suggesting 
a more even sampling of different parts of the belt over very long time spans.} 
Combined with the much larger population of asteroids in the outer main belt, 
this implies that $\sim$21 times more MBAs escape from the outer belt than from the inner belt. 
Dark asteroids are predominant in the outer belt (e.g., the ${\rm d/b}$ ratio for $a>2.82$ au and $D>5$ km 
is $\simeq3.7$; see footnote 3) and the escape statistics is therefore skewed toward dark asteroids as well. 
We find that 2588 dark ($p_{\rm V} \leq 0.1$) and 623 bright ($p_{\rm V}>0.1$) $D>5$ km MBAs escape in 
100 Myr, indicating the overall $\simeq$4.2 preference for dark fugitives (Fig. \ref{escape}c,d). Given 
the incomplete WISE sample, the preference may even be slightly higher if some dark outer belt MBAs are missing 
from the selected sample (Sect 4.1). 

Most escaping asteroids start close to mean motion resonances with Jupiter, such as the 3:1, 8:3, 5:2, 
7:3, 9:4, 11:5, and 2:1 (in the order of increasing semimajor axis). The bodies 
that start sunward from a resonance must have ${\rm d}a/{\rm d}t>0$ to reach the resonance (blue dots 
in Fig. \ref{escape}a,b), whereas the bodies that started beyond a resonance must have ${\rm d}a/{\rm d}t<0$ 
(green dots in Fig. \ref{escape}a,b). A relatively small fraction of the escaping MBAs are members of 
asteroid families (e.g., the Flora family at the inner edge of the asteroid belt, Nysa-Polana complex 
on the sunward side of the 3:1 resonance, Euphrosyne family at $3.1<a<3.2$ au and $25^\circ<i<30^{\circ}$; 
Masiero et al. 2015b). Overall, only $\sim 20$\% of $D>5$ km NEAs are found to be previous members of 
known asteroid families (Sect. 5.3; Nesvorn\'y et al., 2015).
 
\subsection{Population of $D>5$ km NEAs}

There are no NEAs in our simulations at $t=0$. By monitoring the number of model NEAs with time 
we find that it takes some time for the NEA population to build up. Interestingly, dark NEAs reach a 
steady state faster (in $\sim 10$ Myr) in our model than bright NEAs. This makes sense because there 
is a larger flux of dark NEAs from the middle and outer belts, but these NEAs have shorter dynamical 
lifetimes (Table 2, Sect. 5.3), which means that the steady state can be established faster than for 
bright NEAs (for which the situation is the opposite). We find that it takes $\sim 50$ Myr for bright 
NEAs to reach a steady state. After this time, both the dark and bright NEA populations 
start to slowly decline over hundreds of Myr, reflecting the diminishing influx in our model (Sect. 
5.1). Below we report the results from $50<t<400$ Myr. 

\begin{figure}
\epsscale{0.65}
\plotone{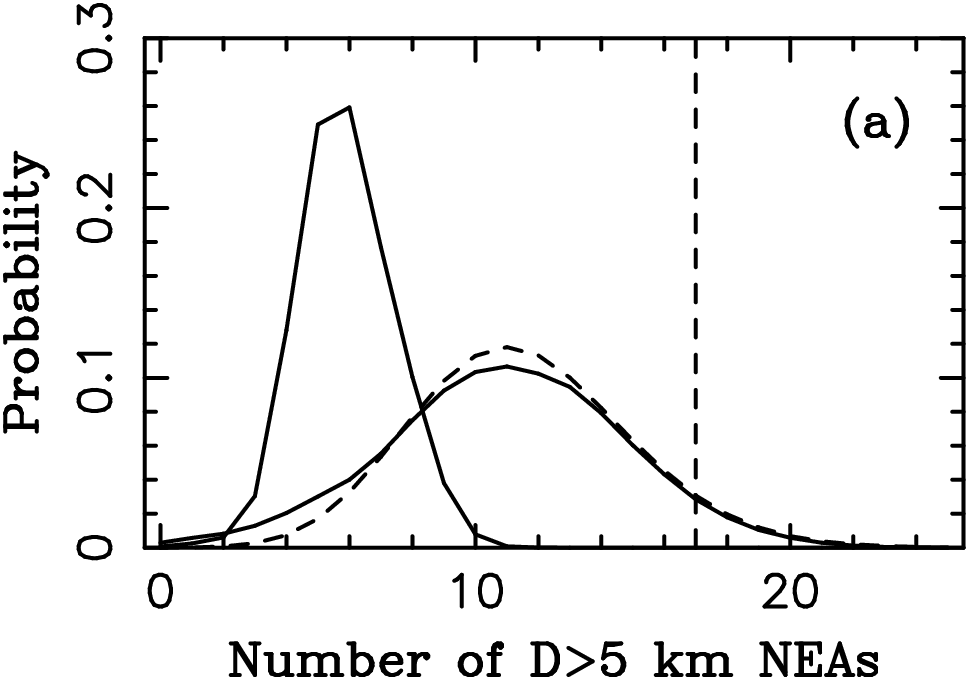}\\[3.mm]
\plotone{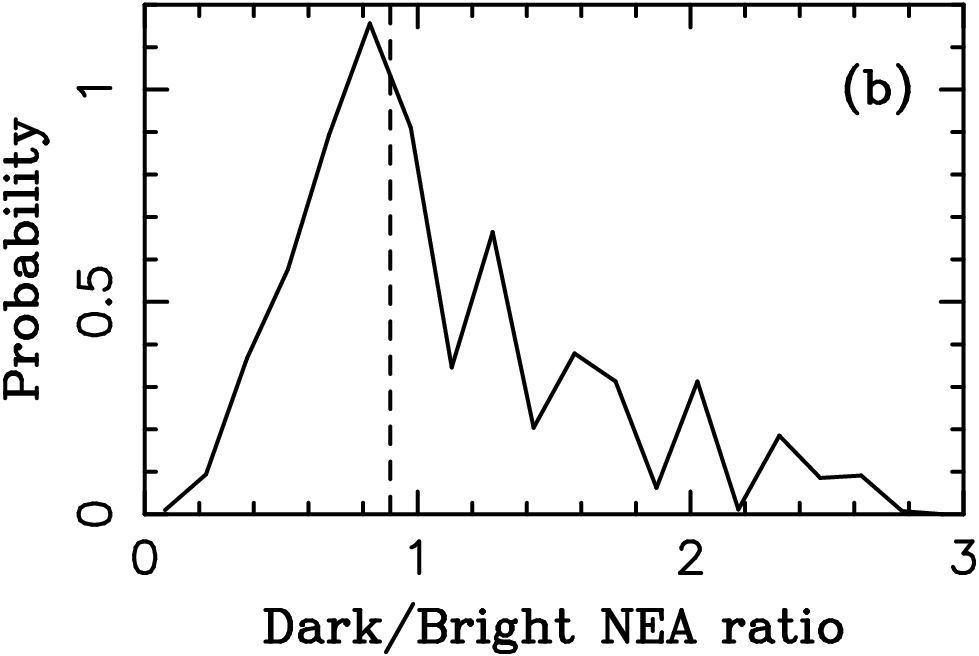}
\caption{Panel (a): The model-derived probability of having a specific number of $D>5$ km NEAs with $q<1.3$ au
and $Q<4.5$ au (solid lines). The two distributions shown here correspond to two different dynamical models 
(see the main text; the left solid line is the random model, the right solid line is the optimized model). 
The dashed line is a reference Poisson distribution. The vertical dashed line shows 
17 NEAs with $D>5$ km from WISE (Table 1). Panel (b): The ratio of dark and bright NEAs with $D>5$ km 
obtained in the model with optimized drift rates. The vertical dashed line is the observed 
ratio: ${\rm d/b} \simeq 0.9$ (9 bright and 8 dark NEAs; Table 1). In both panels, the total probability is 
normalized to 1 and the probability density is shown on the $Y$ axis.}
\label{neapop}
\end{figure}

We find that the number of $D>5$ km NEAs, $n_5$, substantially fluctuates over time. Figure 
\ref{neapop}(a) shows the probability distribution of $n_5$. Here we only consider orbits 
with $q<1.3$ au and $Q<4.5$ au to avoid NEOs with a possible cometary origin (we do not model dormant 
comets here). Two models are plotted. The distribution that peaks for $n_5 \simeq 6$ was obtained 
from all clones included in the integrations. This would correspond to a situation where it is 
equally probable to have maximum negative, zero, and maximum positive Yarkovsky drifts (hereafter 
the {\it random} drift model). The broader distribution that peaks for $n_5 \simeq 11$ was 
obtained by optimizing the Yarkovsky drift. For that, we selected \textit{one clone for each MBA} 
that escaped to a NEA orbit, assuming that such a clone exists for that MBA. If more 
than one clone escaped for the same MBA, we selected one escaping clone at random. The mean 
values are $\langle n_5 \rangle = 6.3$ and 12.6 in the models with the random and optimized 
drifts. Relative to the Poisson distribution $P(n)=e^{-\lambda} \lambda^n/n!$, where $n \geq 0$ and 
the occurrence rate $\lambda=11.5$ (the dashed line in Fig. \ref{neapop}(a)), the optimized 
distribution shows a slight excess of cases with $n_5<7$.

To compare our results with observations, we extracted all $D>5$ km NEAs from the WISE catalog
(Mainzer et al., 2019). The basic information about these NEAs is listed in Table 1. We find that 
there are 17 NEAs with $q<1.3$ au, $Q<4.5$ au, and $D>5$ km. Three of these bodies have $D \simeq 5$~km 
(Table 1) but a rather large diameter uncertainty, and in reality can be smaller than 5 km.
It is also possible that we missed some $D>5$ km NEAs either because they were not observed by 
WISE or because their diameter was sub-estimated from thermal modeling. For reference, Nugent 
et al. (2016) reported the diameter errors from WISE have $\sim 20$\% (1$\sigma$) uncertainties.

The NEA model with random Yarkovsky drifts is clearly incompatible with observations (Fig. 
\ref{neapop}(a)). In that model, the probability to have $n_5 \geq 17$ is $<10^{-5}$. The model 
with optimized drifts fares better: $n_5 \geq 17$ is expected with the $\simeq$11\% probability. 
Still, the current population of NEAs is larger than the long-term average, which may give support 
to the possibility that the present impact flux on the terrestrial worlds has increased 
$\sim 300$--400 Myr ago (e.g., Culler et al., 2000; Mazrouei et al., 2019; also see Kirchoff et al., 
2021). The optimized model implies that MBAs are drifting toward resonances that lead to their ultimate 
escape from the main belt (and not away from them). This makes sense because if MBAs would be drifting 
away from resonances, they would need to start in the resonances, and would not exist in the 
first place (would be removed in the past). There are several potential caveats to this. For 
example, MBAs may jump over weaker resonances and may appear as drifting away from them at 
the present time. 

The observed ratio of dark/bright NEAs with $D>5$ km is ${\rm d}/{\rm b} = 8/9 \simeq 0.9$ (Table 1). 
Figure \ref{neapop}(b) shows the ${\rm d}/{\rm b}$ distribution obtained in the model with optimized drifts 
(the distribution with random drifts is similar). As the number of large NEAs changes with time, the
ratio of dark/bright NEAs changes as well. This is a consequence of the stochastic delivery process.
Thus, depending on the considered time, it can be as low as $<0.5$ ($\sim10$\% probability) or as high 
as $>2$ ($\sim5$\% probability). The most likely values, however, are intermediate: the ratio distribution 
peaks near ${\rm d}/{\rm b}=0.8$ and the mean value is $\langle {\rm d}/{\rm b} \rangle = 1.3$. There 
is thus a relatively good agreement with observations. With that said, however, it needs to be mentioned  
that the observations obtained at the current epoch and are not particularly constraining for the 
long-term average. For comparison, Morbidelli et al. (2020) estimated ${\rm d}/{\rm b}=1.3$ for small 
NEOs from calibrating the many-source model of G18 on the NEOWISE observations (Mainzer et al. 2019).

\subsection{NEA Sources}

We now consider the source regions of $D>5$ km NEAs. For that, we select all NEAs produced in 
our simulation for $50<t<400$ Myr and track these bodies back to their starting orbits. We find that the 
inner, middle and outer belts contribute by 52\%, 35\% and 13\%, respectively. The percentages quoted here 
were obtained in the model with optimized drifts but the values for random drifts are similar (Table 3). 
We therefore estimate that MBAs with $a<2.5$ au and $a>2.5$ au supply roughly the same number of $D>5$ km 
NEAs. This implies a much larger contribution of the middle/outer main belt than found previously 
for small NEAs from the many-source models (B02, G18; see Sect. 6.1 for a discussion).  

\begin{figure}
\epsscale{0.65}
\plotone{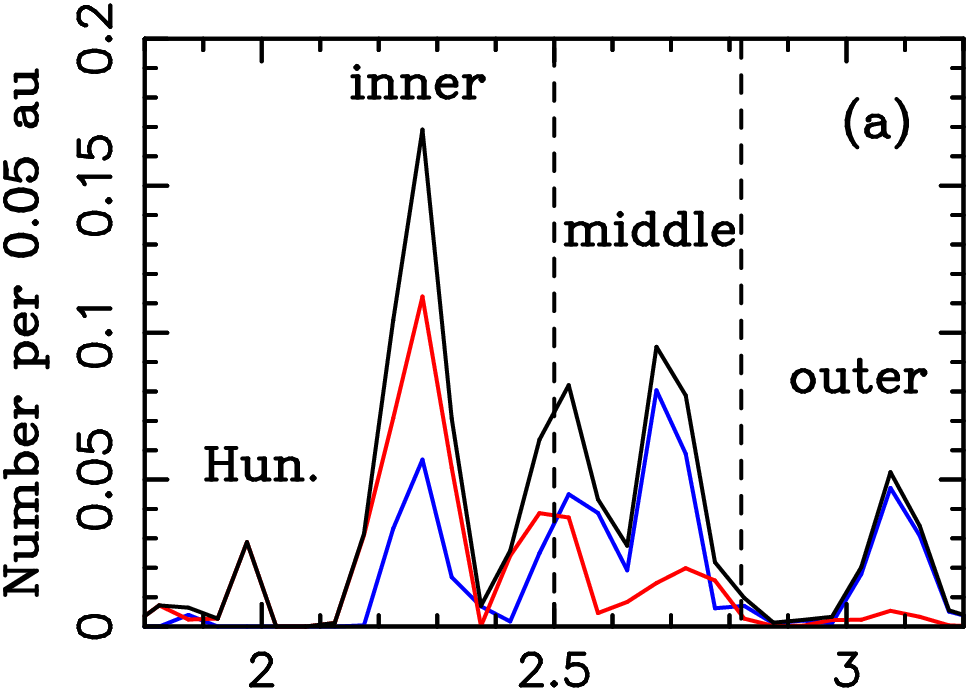}\\[3.mm]
\plotone{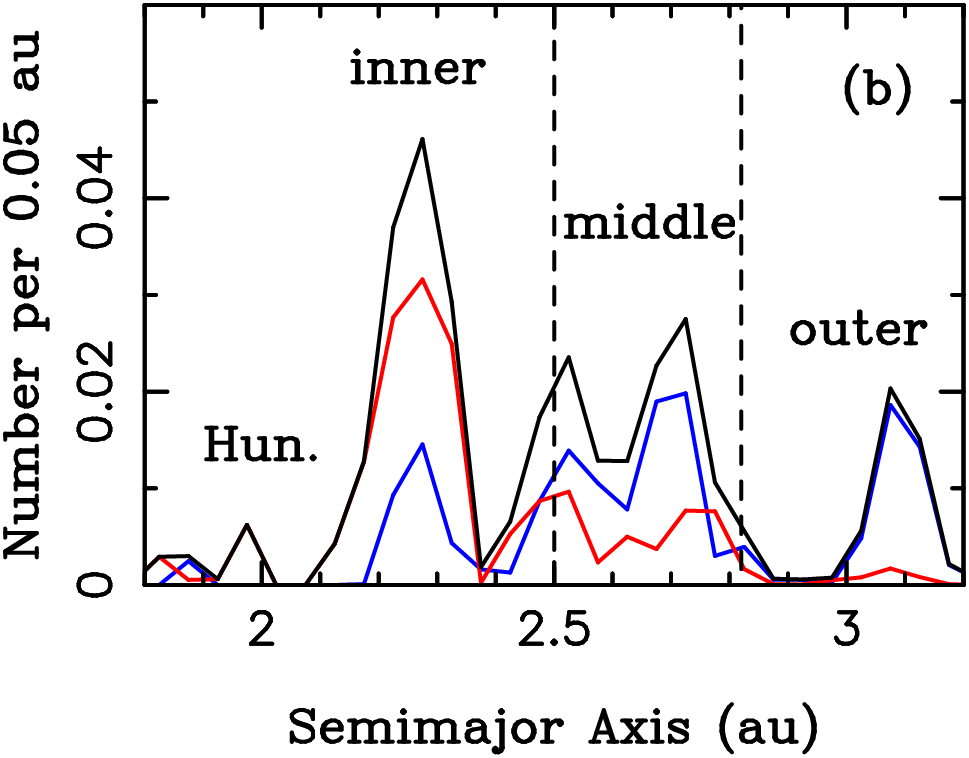}
\caption{The contribution of MBAs to NEAs is shown here as a function of MBA's starting semimajor axis 
(black solid lines). The blue/red solid lines show the main belt contribution to dark/bright $D>5$ km NEAs,
respectively. The inner belt is the main source of bright NEAs and the middle belt is the main source of dark NEAs.  
The optimized drift model is shown in panel (a), random drift model in panel (b). Label ``Hun.'' stands for 
Hungarias.}
\label{source}
\end{figure}

For example, B02 determined from the many-source model that the combined contribution of $\nu_6$, IMCs and 
3:1 is 84\%, leaving only 16\% for the outer belt and JFCs.  We find that the contribution of MBAs with 
$a<2.55$ au, which includes $\nu_6$, IMCs and 3:1, is significantly lower, 61\%, leaving a much larger 
contribution to the middle/outer belt (39\%; Fig. \ref{source}). It is more difficult to compare our 
results to G18, because the seven sources used in that work included extended `complexes' of escape 
routes near major resonances. To estimate the inner belt contribution from G18, we put together the 
contributions of Hungarias, Phoaceas, the $\nu_6$ complex, and the 3:1 complex. This gives 83\%, which 
is similar to the estimate of B02, and leaves only 17\% for the middle/outer belt and comets. Again, 
this is much lower than our 39\%.

The differences discussed above may have several different interpretations: (i) The contribution of 
different sources is size-dependent; the many-source models were calibrated on $D \lesssim 1$ km NEOs, 
whereas here we have $D>5$ km. (ii) The many-source models, given the methodology caveats discussed in 
Sect. 2, may fail to properly weight in the contribution of MBAs with $a>2.5$ au. (iii) The contribution 
of different sources is time variable and the present NEO distribution -- as characterized from 
many-source models -- is not representative for the long term average. See Section 6.1 for a discussion. 
For Hungarias and Phoaceas, we find the 4.5\% and 9.1\% contributions, respectively, whereas G18 
reported 5.6\% and 2.7\%. The reason behind the disagreement for Phoaceas is unclear.

In our single-source model, the contribution of MBAs with $a<2.5$ au and $a>2.5$ au to large NEAs is 
similar (52\% and 48\%, respectively; Table 3). This has interesting consequences for how the main belt 
supplies dark/bright $D>5$ km NEAs. The middle belt is the main contributor of dark NEAs (50\%; Table 4, 
Fig. \ref{source}) and the inner belt is the main contributor of bright NEAs (76\%). In B02 and G18, 
instead, where over 80\% of NEOs originate in the inner belt, most dark NEAs also come from the inner 
belt (G18, Morbidelli et al. 2020). These differences may be related to some of the issues mentioned 
above. For example, they could reflect the size-dependent nature of the delivery process.  

The contribution of asteroid families to $D>5$ km NEAs is minor. To determine this contribution,
we link the $D>5$ km MBAs to the family catalog from Nesvorn\'y et al. (2015). We find that only 20\% of
$D>5$ km NEAs in a steady state are expected to come from known families. The remaining 80\% come from 
background MBAs. This estimate is consistent with Vokrouhlick\'y et al. (2017), who found that the 
Flora family, which is optimally placed near the $\nu_6$ resonance to produce NEAs, contributes by only 
3.5--5\% to the $D>1$ km NEA population at the present time. The individual contribution of any other asteroid family to 
the present-day NEAs is low ($<1$\% for $D>5$ km). The family contributions to smaller NEAs could be 
more significant because the collisional families generally have a steep size distribution for $D 
\lesssim 5$ km (Masiero et al., 2013, 2015b).

Dynamical lifetimes of NEAs are listed in Table 2. We find that the mean dynamical lifetime of $D>5$ km 
NEAs is $\langle t_5 \rangle = 1.2$ Myr, which is shorter than the $\sim 4$ Myr estimate from B02. This 
probably reflects a greater contribution of the middle/outer belt in our model. We identify a clear 
trend with the dynamical lifetime decreasing with the radial distance of a source. For example, MBAs 
evolving from the outer belt only spend, on average, $\simeq 0.5$ Myr on NEA orbits. Compared to 
Gladman et al. (1997), we find a good agreement for the 3:1 and 5:2 resonances ($\langle t_5 \rangle = 2.9$ 
and 0.7 Myr, respectively). For 8:3, which is an important source region of dark NEAs in the middle belt 
(Fig. \ref{source}), we estimate $\langle t_5 \rangle = 2.3$ Myr.  

\subsection{Orbital Distribution of Large NEAs}

\begin{figure}
\epsscale{0.7}
\plotone{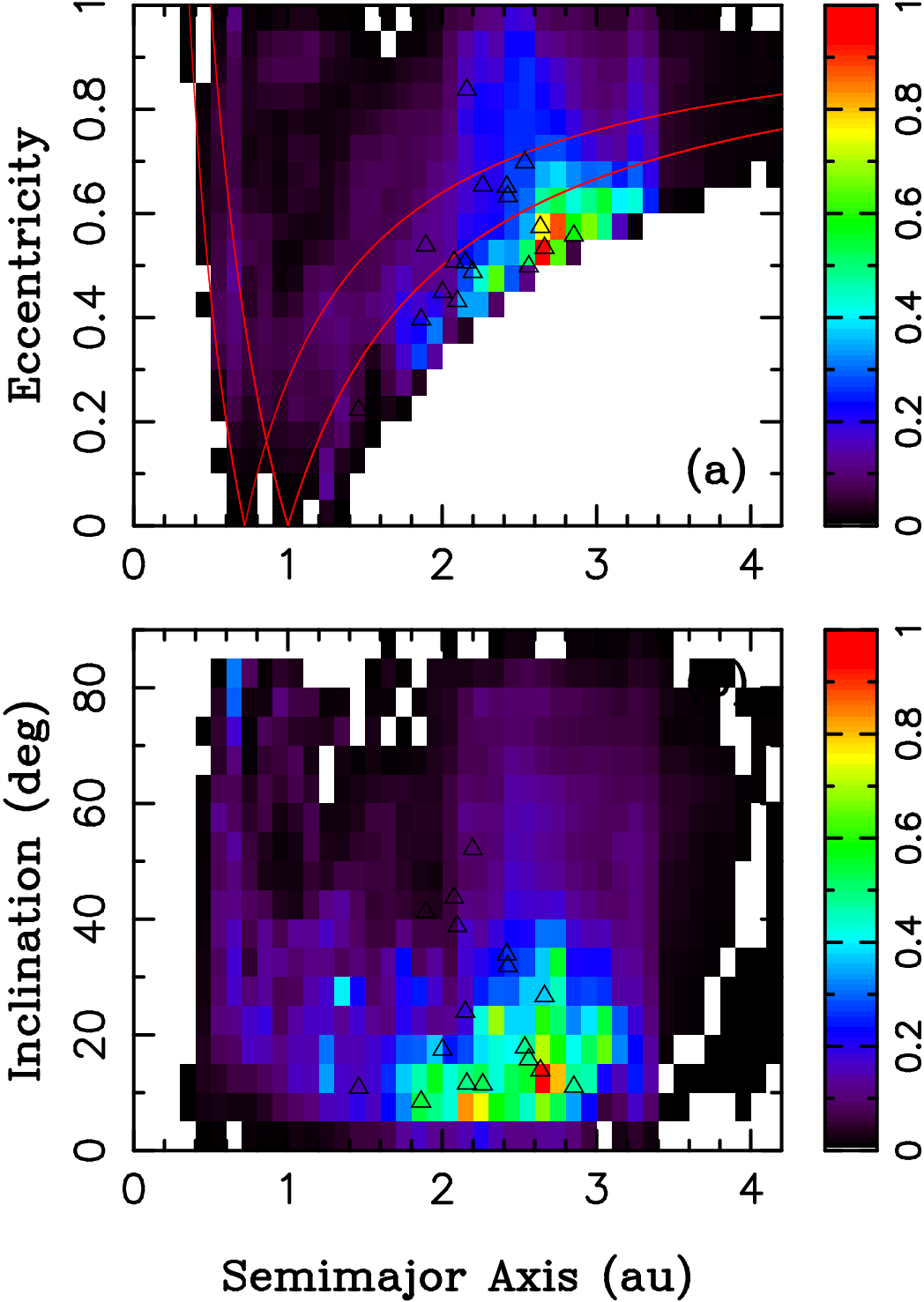}
\caption{The orbital distribution of $D>5$ km NEAs obtained in our model with optimized drift rates. 
The binned residence times from our simulations are normalized to one and plotted here in the $(a,e)$ and 
$(a,i)$ projections. The color scale, which corresponds to the binned probability, appears on the 
right (warm colors indicate higher probability). The triangles show 17 known NEAs with $D>5$ km 
(Table 1).}
\label{resid}
\end{figure}

\begin{figure}
\epsscale{0.49}
\plotone{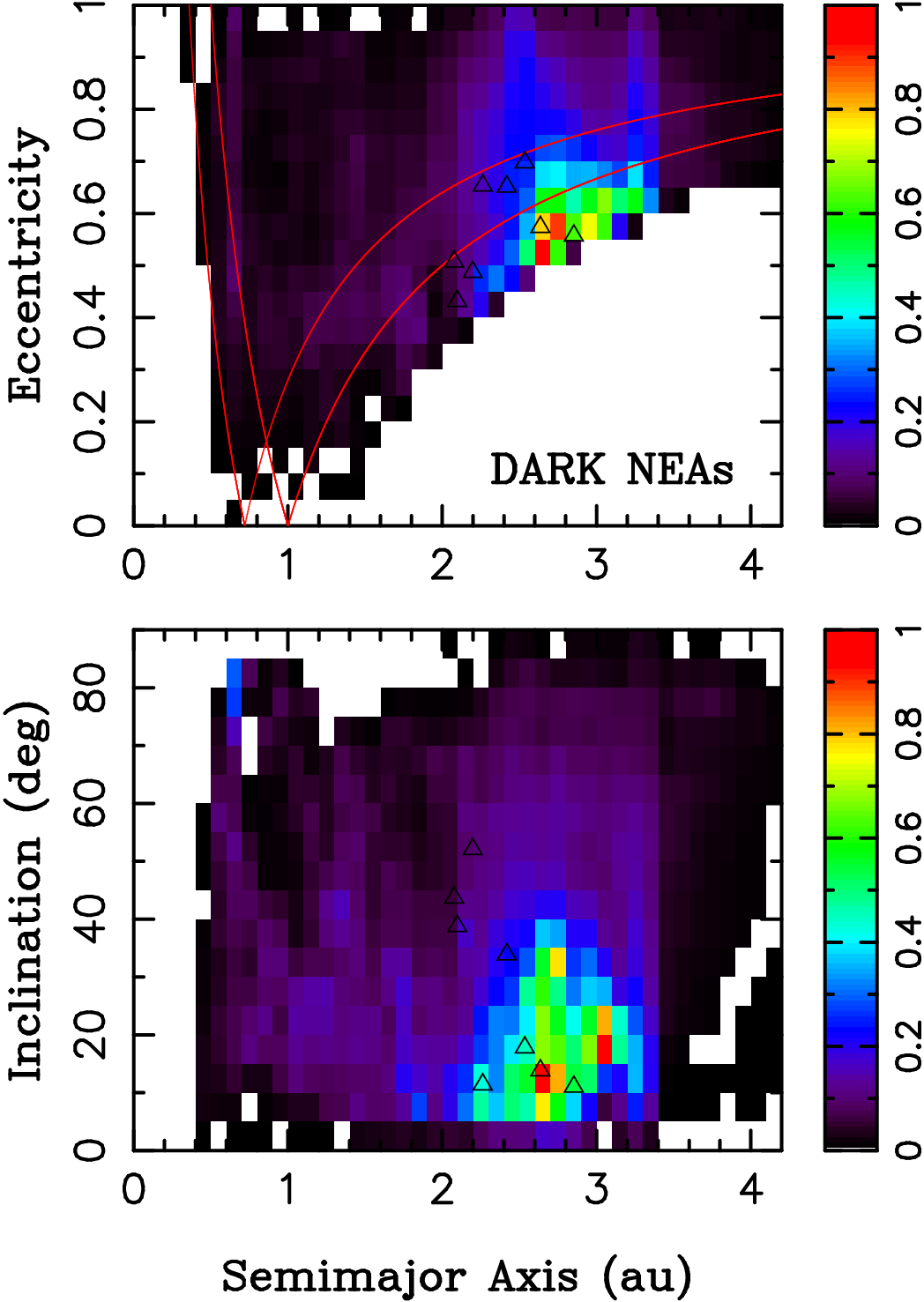}
\plotone{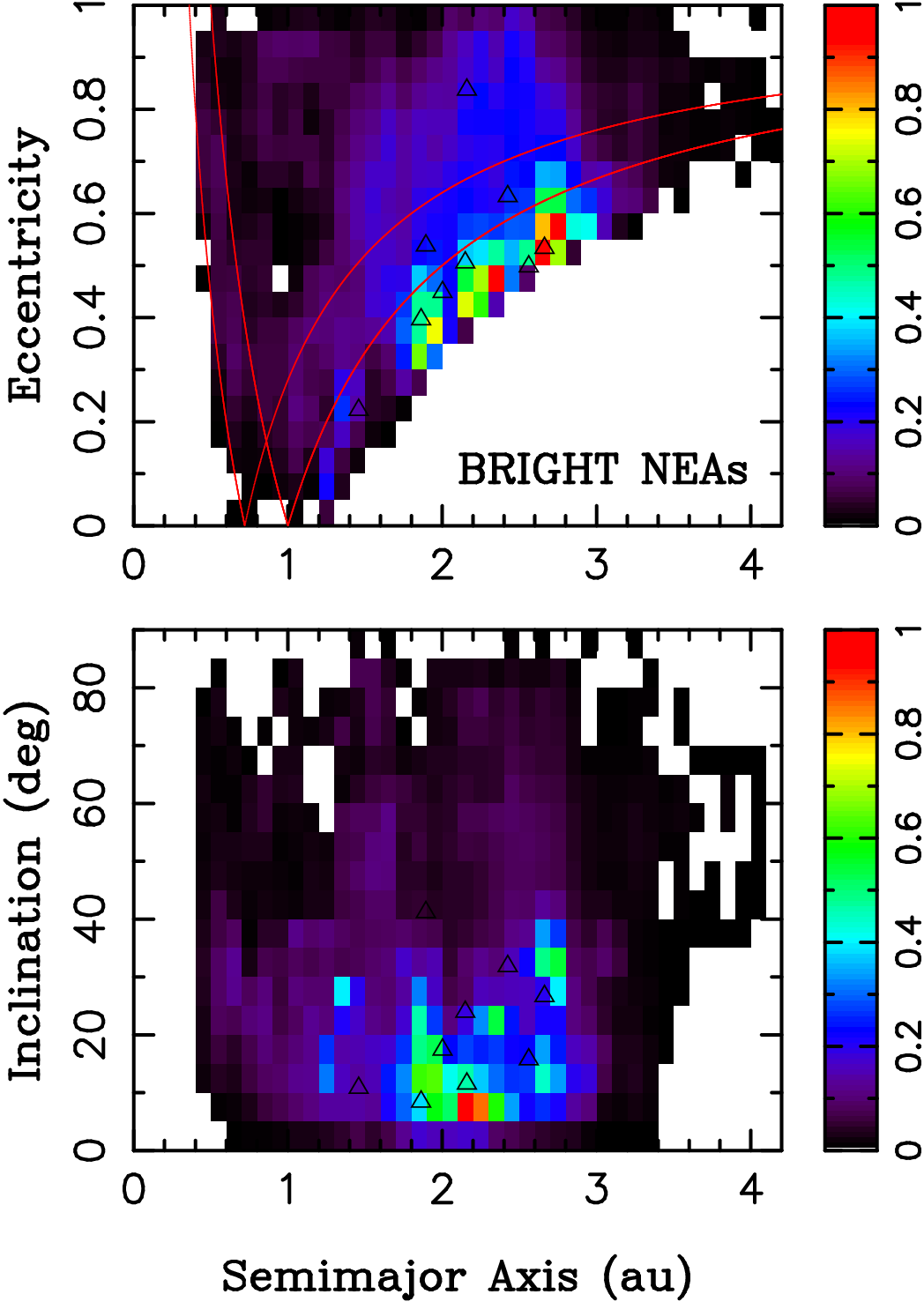}
\caption{The orbital distribution of $D>5$ km NEAs obtained in our model: $p_{\rm V} \leq 0.1$ on the left and
$p_{\rm V} > 0.1$ on the right. The two distributions are slightly different: the highest probability for dark 
NEAs occurs for orbits $2.6<a<3.2$ au, whereas bright NEAs are expected to populate a wide radial range, 
$1.8<a<2.8$ au. The triangles show 8 known $D>5$ km NEAs with $p_{\rm V} \leq 0.1$ (left panels) and 9 known 
$D>5$ km NEAs with $p_{\rm V} > 0.1$ (right panels).}
\label{resid2}
\end{figure}

Our single-source model can be used to determine the steady-state orbital distribution of large NEAs. 
Here, there is no need for observational calibration of different sources. The orbital distribution of NEAs 
is uniquely determined by the main belt structure and our (simple) physical model for the radiation forces. 
It represents a testable model prediction. Figure \ref{resid} shows the orbital distribution of $D>5$ km 
NEAs obtained in the model. The distribution peaks at $2<a<3$ au, $1<q<1.3$ au and $5^\circ<i<30^\circ$. 
Compared to B02 and G18, where similar plots were published for smaller NEAs, there are more orbits with 
$a>2.5$ au (and fewer orbits with $a<2.5$ au). Dark $D>5$ km NEAs are primarily responsible for this shift 
(Fig. \ref{resid2}, left panels). The middle/outer belt ($a>2.5$ au) supplies the majority of dark 
NEAs in our model (71\% total contribution; Table 4), and this leads to a distribution that is skewed 
toward $a>2.5$ au. Bright $D>5$ km NEAs are distributed more equally between 2 and 3 au (Fig. 
\ref{resid2}, right panels).

\begin{figure}
\epsscale{0.3198}
\plotone{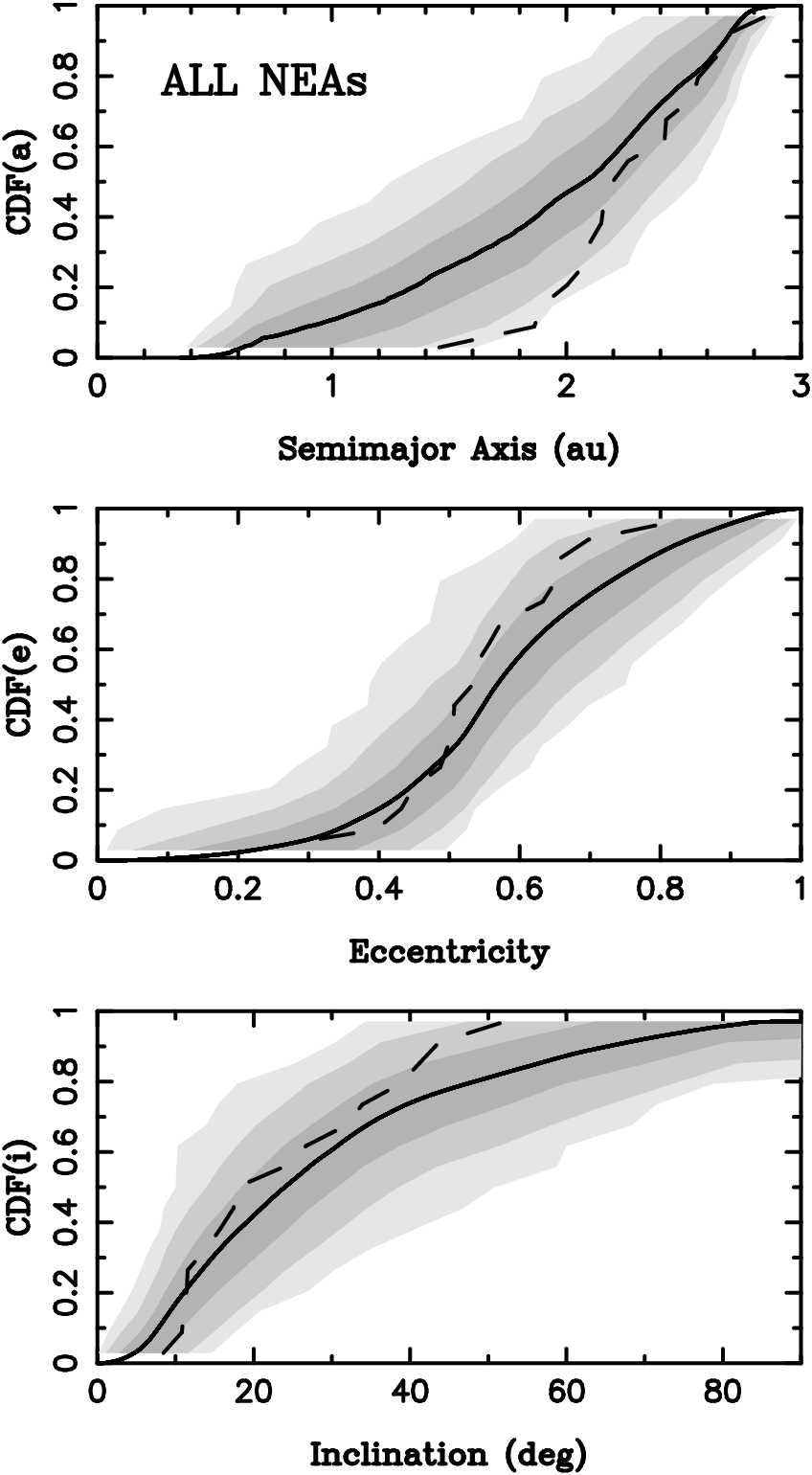}
\epsscale{0.3}
\plotone{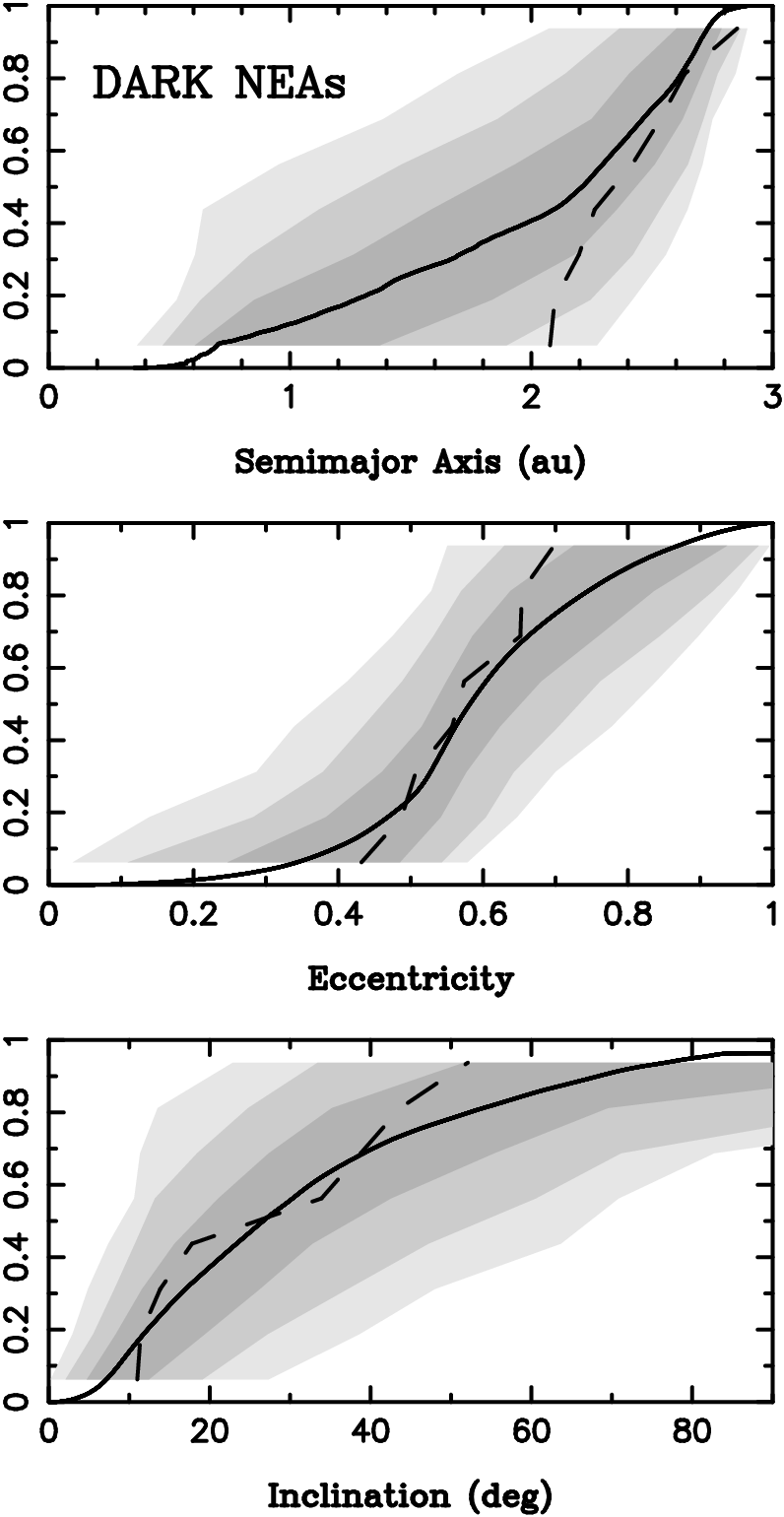}\hspace{1.mm}
\plotone{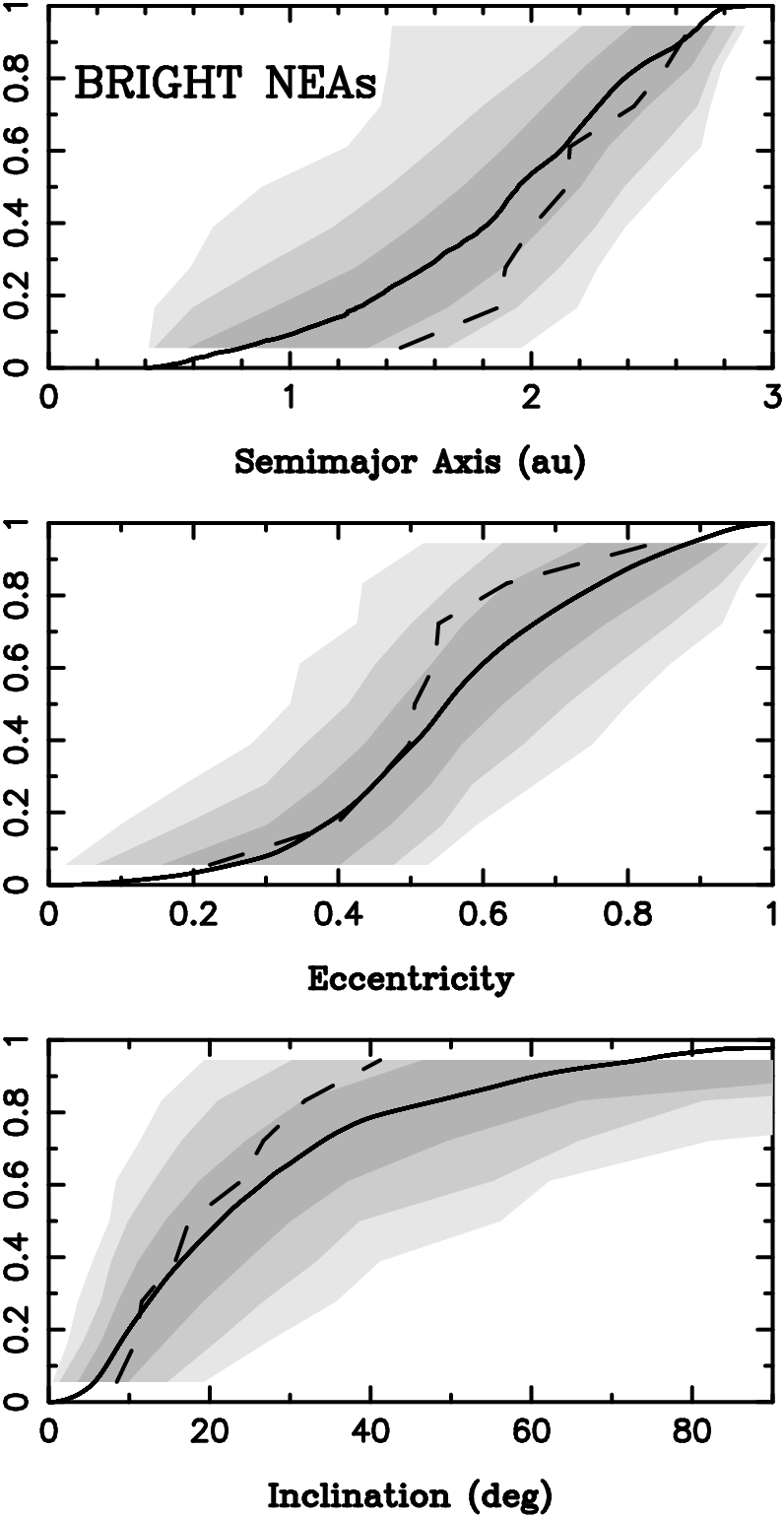}
\caption{The cumulative distribution functions (CDFs): all NEAs (left panels), dark NEAs with $p_{\rm V} \le 0.1$ 
(middle panels), and bright NEAs with $p_{\rm V} > 0.1$ (right panels). The solid and dashed lines show the 
optimized model and observed distributions for $D>5$ km NEAs, respectively. The shaded ares are an expression of the 
statistical uncertainty: 1$\sigma$ (68.2\%, dark gray), 2$\sigma$ (95.4\%, grey), and 3$\sigma$ (99.7\%, 
light grey). To define these areas we sub-sampled the model results to the number of real NEAs in each 
category (17 for all, 8 for dark, 9 for bright), generated 1000 sub-sampled distributions, and defined each 
envelope to contain the appropriate percentage of trials.}
\label{cumul}
\end{figure}

We compare the model distributions with observations in more detail in Fig. \ref{cumul}. Unfortunately, 
the number of $D>5$ km NEAs is statistically small and the comparison is not particularly constraining. 
A potential problem is identified in the semimajor axis distribution (or, equivalently, the perihelion 
distance distribution). This is particularly clear for dark NEAs, where all 8 known NEAs with $D>5$ km 
and $p_{\rm V} \leq 0.1$ have $a>2$ au, whereas our model suggests that $\sim$40\% of dark/large NEAs 
should have $a<2$~au. Statistically, this represents odds of $0.6^8 \simeq 1.7$\% (i.e., between 2$\sigma$ 
and 3$\sigma$). When the dark and bright NEAs are combined, the statistics improves (left panels in Fig. 
\ref{cumul}), but the semimajor axis difference remains below 3$\sigma$. The small number statistics 
has a heavy influence on this comparison. In addition, our simulations ignore Mercury (Sect. 4.3), 
and we are therefore not fully confident that the orbital distribution of model NEAs with low perihelion 
distances is correct. 

\begin{figure}
\epsscale{0.7}
\plotone{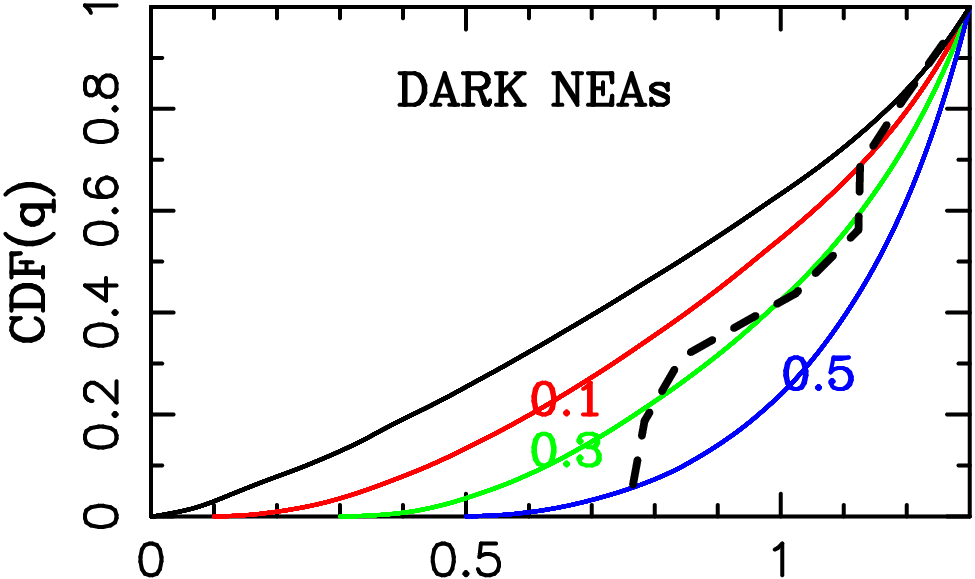}\\[3.mm]
\plotone{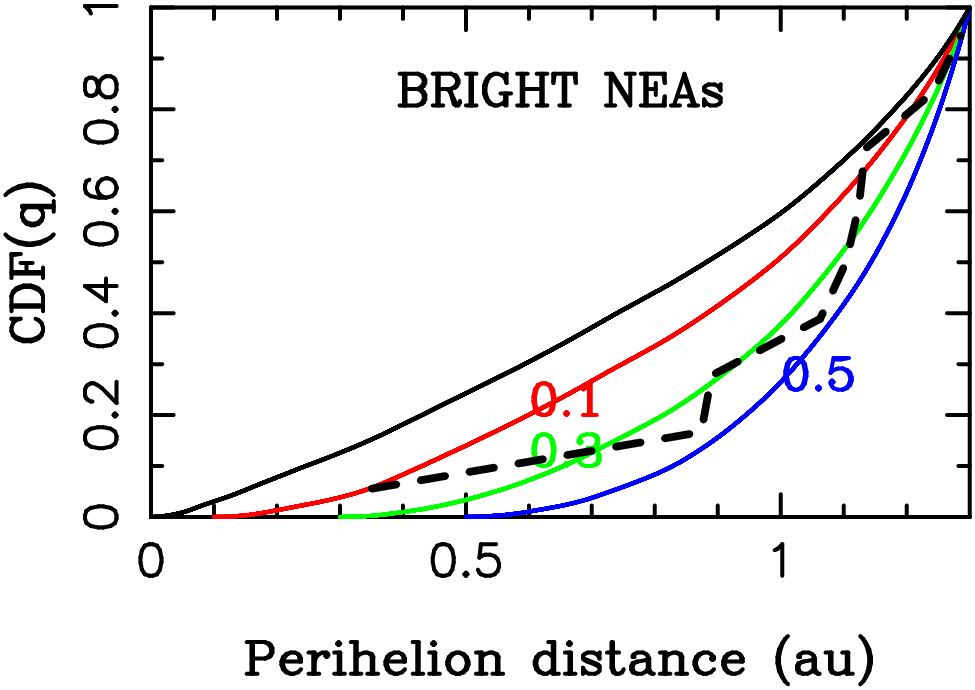}
\caption{The perihelion distribution of $D>5$ km NEAs: dark ($p_{\rm V} \leq 0.1$) in the top panel 
and bright ($p_{\rm V} > 0.1$) in the bottom panel. The solid lines correspond to different models. 
The black line is the nominal model from Fig. \ref{cumul}; the large statistical uncertainty is not 
shown here. The color lines correspond to the models where NEAs evolving to $q<q^*$ were removed: 
$q^*=0.1$ au (red), $q^*=0.3$ au (green), and $q^*=0.5$ au (blue). The dashed line is the perihelion 
distribution of known $D>5$ km NEAs. Only one of 17 known $D>5$ km NEAs has $q<0.7$~au (2212 
Hephaistos with $p_{\rm V}=0.16$ and $q=0.35$ au).}
\label{peri}
\end{figure}

Despite these important caveats, we investigated models where NEAs reaching perihelion distance $q<q^*$, 
where $q^*$ is a free parameter, are removed. This is motivated by the possibility that thermal stresses 
close to the Sun could break up mechanically weak NEAs (Delbo et al. 2014, Granvik et al. 2016). A prime 
example of this is 3200 Phaethon with $q=0.14$ au, the parent body of the Geminid meteoroid stream, which 
episodically loses mass at an average rate of $\sim700$ kg s$^{-1}$ (Jewitt et al. 2019). The best match 
to the observed distribution is obtained for $0.1 \lesssim q^* \lesssim 0.3$ au (Fig. \ref{peri}). This 
could suggest that large NEAs (gradually?) disintegrate when their orbital perihelion drops below 
0.1-0.3 au. For comparison, Granvik et al. (2016) suggested that $D\lesssim$ 1 km NEAs super-catastrophically 
disrupt for $q \lesssim 0.1$ au.

\subsection{Impact Flux from Large NEAs}
  
\begin{figure}
\epsscale{0.49}
\plotone{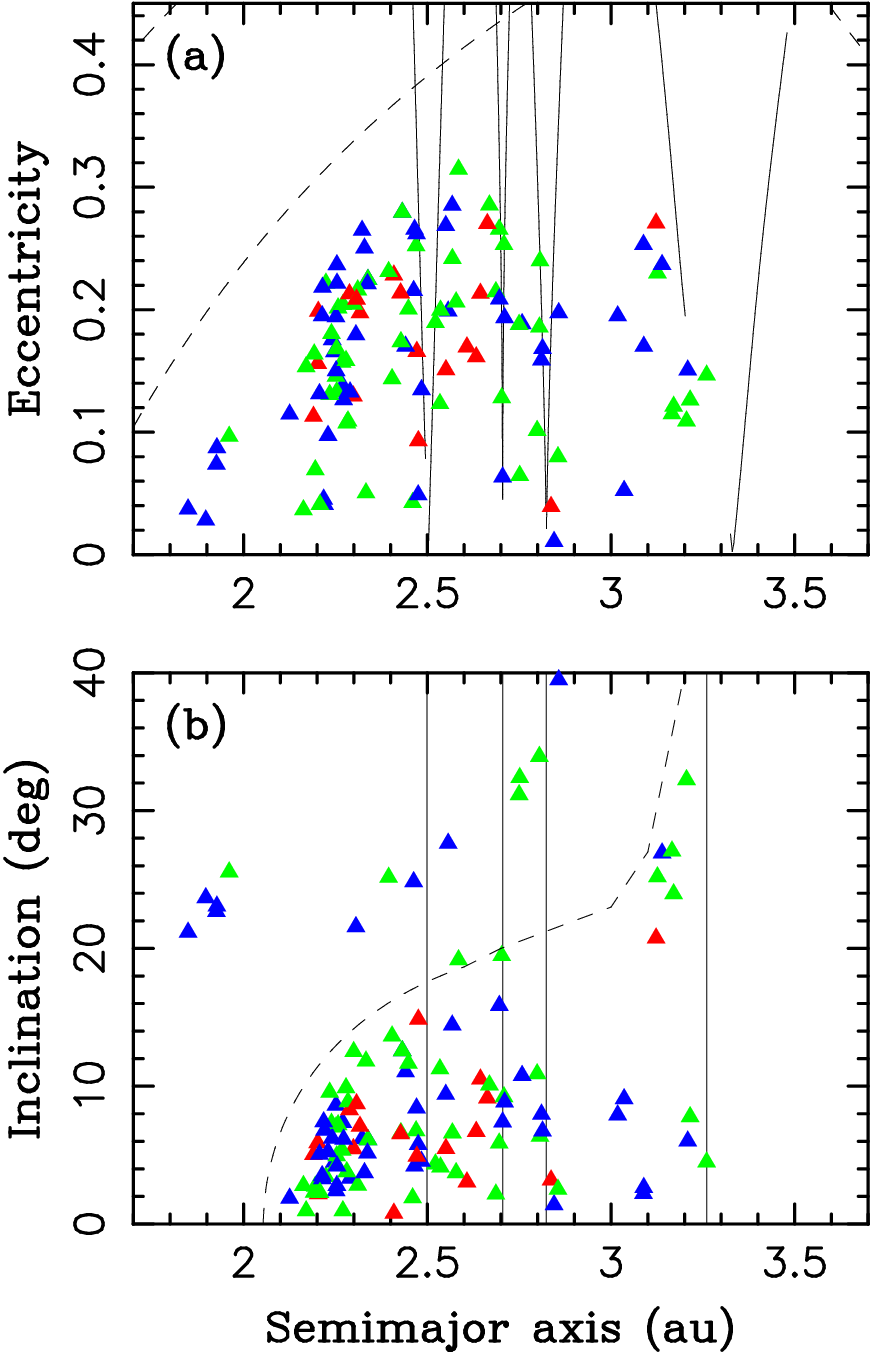}
\plotone{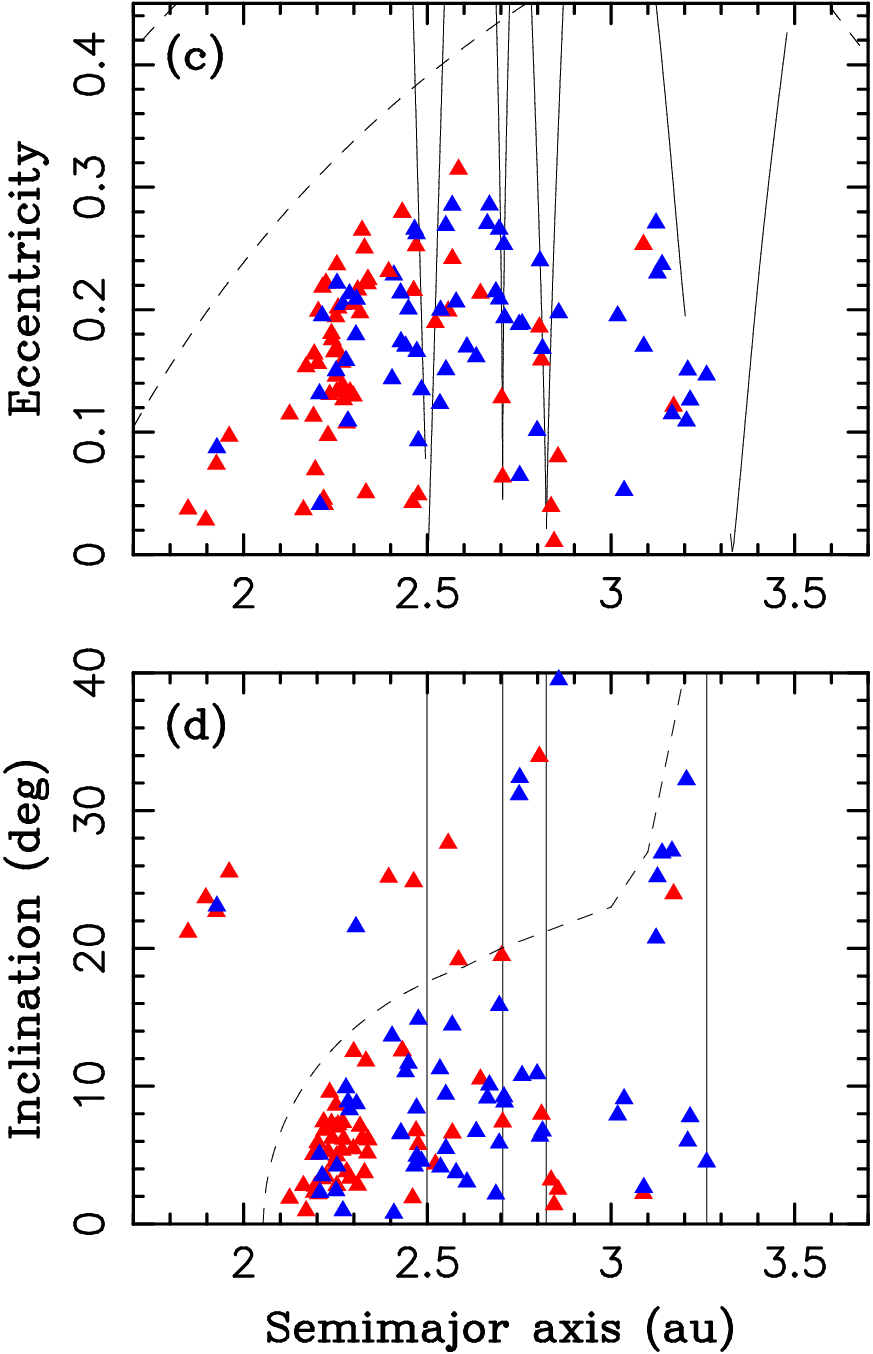}
\caption{Panels (a) and (b): The initial orbits of $D>5$ km asteroids that ended up impacting on the terrestrial
planets: Venus (green), Earth (blue) and Mars (red). Panels (c) and (d): The albedo of impacting asteroids: 
$p_{\rm V} \leq 0.1$ (blue) and $p_{\rm V} > 0.1$ (red).}
\label{hit}
\end{figure}

In total, including all clones, 119 impacts of $D>5$ km asteroids were recorded on Venus (52 impacts), 
Earth (49) and Mars (18) in 1 Gyr (Fig. \ref {hit}). Thus, in the random drift model, we infer $49/3 
\simeq 16$ $D>5$ km asteroid impacts on the Earth in 1 Gyr. The impact rate is about twice as high in the optimized 
drift model. We thus estimate $\sim$16--32 $D>5$ km asteroid impacts on the Earth in 1 Gyr, and the average 
time between impacts $\sim 30$--60 Myr. The impact flux on Venus is similar, and Mars receives roughly 37\% 
of the terrestrial flux (i.e., the Earth-to-Mars impact flux ratio for $D>5$ km asteroids is $\sim2.7$). 

Previous studies of small NEOs estimated that the average impact probability for one object 
in the NEO population is $p_{\rm i}=1.5\times10^{-3}$ Myr$^{-1}$ (Stuart 2001, Harris \& D'Abramo 2014) or 
$p_{\rm i}=1.3\times10^{-3}$ Myr$^{-1}$ (Morbidelli et al. 2020). When these probabilities are multiplied 
by the number of known $D>5$ km NEAs (17; Table 1), we compute $\simeq$22--26 impacts on the Earth in 1 
Gyr, which would be consistent with our estimate from the recorded impacts. It is not clear, however, 
whether this comparison is strictly correct because of different definitions of the NEA/NEO target 
regions in different works (e.g., $a<4.2$ au in G18 and Morbidelli et al. (2020), but $Q<4.5$ au here). 
If NEOs with $a<4.2$ au and $Q>4.5$ au were included here, the number of impacts estimated from $p_{\rm i}$ 
would be higher. 

For comparison, Johnson et al. (2016), adopting $p_{\rm i}=1.5\times10^{-3}$ Myr$^{-1}$ and using absolute 
magnitude as a proxy for size, estimated $\sim$50 impacts of $D>5$ km asteroids on the Earth in 1 Gyr, which 
is 1.6-3.1 times higher than our value. Nesvorn\'y et al. (2017) and Nesvorn\'y \& Roig (2018) estimated 
that only $\sim 1$ impact of a $D>10$ km asteroid should occur on the Earth over 1~Gyr. Here we find 
$\sim 2$ such impacts in the random drift model. Ignoring the Yarkovsky effect and rescaling from $D>10$ km 
to $D>5$ km from the MBA-inferred ratio of $\sim$5.2 (Sect. 5.1), we would infer only $\sim 10$ impacts of 
$D>5$ km asteroids on the Earth per 1 Gyr. Our best estimate is $\sim1.6$--3.2 times higher than that. 
This means that the Yarkovsky effects kicks in for $D < 10$ km and generates more $D>5$ km NEAs than what 
would be expected from a simple scaling based on the size distribution of MBAs. 

We see some variability when the impact statistics obtained in our model is sliced in time (e.g., 
the number of impacts in a 100-Myr interval fluctuates by a factor of $\sim 2$), but no long-term 
trends (there are 56 impacts in $0<t<0.5$ Gyr and 63 impacts in  $0.5<t<1$ Gyr). Here we therefore 
report the results for the full 1-Gyr interval, where we have the best statistics. There were 72 
impacts (60\%) from bodies starting with $a<2.5$ au and 47 impacts (40\%) from $a>2.5$ au. The inner 
belt thus supplies somewhat more impactors on the terrestrial worlds, $\sim 1.5$ times more, than the 
middle/outer belt. For comparison, G18 reported that $\sim$80\% of terrestrial impactors with 
absolute magnitudes $17<H<22$ ($0.17<D<1.7$ km for $p_{\rm V}=0.1$) come from the $\nu_6$ resonance at 
the inner edge of the main belt. Relative to this, the middle/outer belt has a much larger 
importance as the source of {\it large} terrestrial impactors.

We find that 47\% of large terrestrial impactors are dark ($p_{\rm V} \leq 0.1$) and 53\% are bright 
($p_{\rm V} > 0.1$), suggesting a roughly equal split (Table 5). About 59\% of dark impactors 
come from the middle/outer belt ($a>2.5$ au).\footnote{Gladman et al. (1997) and Bottke et al. (2006b) 
found that the 2:1 resonance produces NEAs with a 0.02\% impact probability on the Earth. Here we 
obtain a 0.07\% Earth-impact probability for large asteroids evolving from the outer belt ($a>2.82$ au), 
which is 3.5 times higher than the estimate quoted above. The 2:1 resonance results should therefore not 
be used as an indicator of the impact flux from the outer belt, at least not for the large impactors.}
In contrast, most {\it small} and dark impactors on the Earth start in the inner belt (B02; G18), suggesting 
a size-dependent sampling of the main belt. 

About 78\% of bright $D>5$ km terrestrial impactors originate in the inner belt ($a<2.5$ au). Of these, 
11 impactors (18\%) were previous members of the Flora family. In total, the asteroid families contribute 
by only 23\% to impacts of $D>5$ km asteroids on the terrestrial worlds (this estimate does not account for 
impactors from the family halos; Nesvorn\'y et al., 2015). Other notable families with more than one recorded 
impact are: Euphrosyne (3 impacts representing 5\% of dark impactors; cf. Masiero et al. 2015a), Koronis 
(3 impacts) and Dora (2 impacts). This implies that a great majority of large terrestrial impactors are 
not previous members of the asteroid families -- they are background MBAs.

The statistics of impacts discussed above would be affected if NEAs break up when they evolve too close 
to the Sun (Sect. 5.4). We investigated this effect and found that the number of impacts on the Earth drops to 
67\% for $q^*=0.1$ au and 51\% for $q^*=0.3$ au (both percentages quoted with respect to the nominal 100\% 
impact flux without the NEA removal). The effect is stronger for Venus (48\% for $q^*=0.1$ and 31\% for $q^*=0.3$ au) 
and weaker for Mars (only a 5\% reduction of the number of impacts for both $q^*=0.1$ and 0.3 au). 
We did not find any obvious dependence of the removal effect on the NEA source (inner vs. outer belt) or 
type (dark vs. bright).  

\begin{figure}
\epsscale{0.7}
\plotone{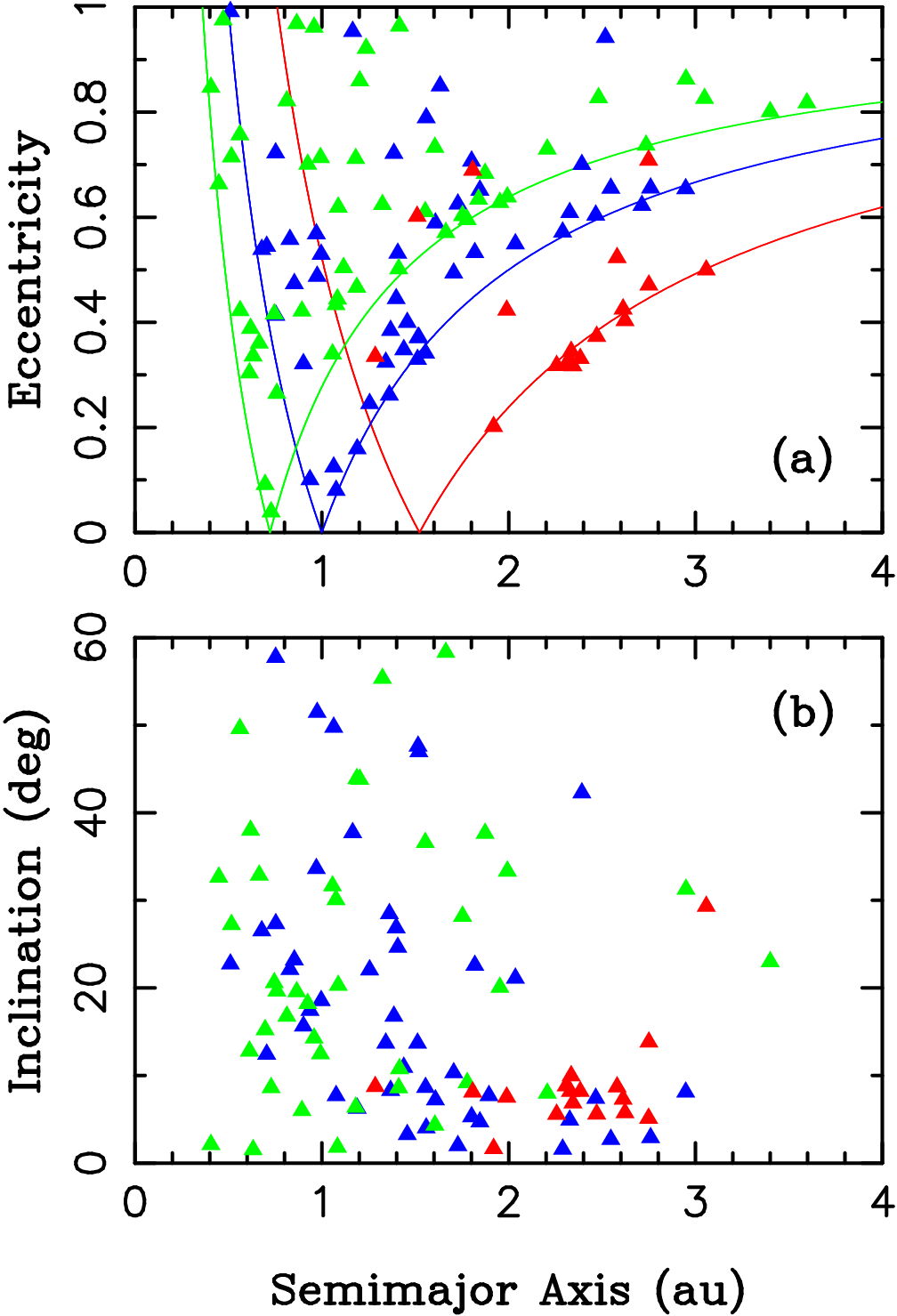}
\caption{The pre-impact orbits of $D>5$ km impactors on the terrestrial worlds: Venus (green), Earth (blue), and Mars 
(red). The orbits are shown when impactors enter 3 planetary Hill radii.
The solid lines in panel (a) show the Tisserand tails of each planet ($q=a_{\rm pl}$ and $Q=a_{\rm pl}$, where 
$q$ and $Q$ are the asteroid's perihelion and aphelion distances, and $a_{\rm pl}$ is the planet's semimajor axis).}
\label{info}
\end{figure}

The pre-impact orbits are shown in Fig. \ref{info}. These are the orbits of $D>5$ km impactors on the 
terrestrial planets just before their final dive toward an impact (the orbits are computed at 3  
Hill radii from a planet). The Venus and Earth impactors have a 
wide orbital distribution with large eccentricities and large inclinations. Most impacts on the Earth 
happen from orbits with $0.8<q<1$ au; that is where the impact probabilities are the highest (Bottke et 
al. 2020). For Mars, 11 out of 18 pre-impact orbits (61\%) have $a>2$ au, $q>1.4$ au and $i<10^\circ$. 
Most $D>5$ km Mars impactors reach the Mars-crossing orbits via weak resonances (Migliorini et al., 
1998), and apparently impact before their perihelion distance evolves too much (Fig. \ref{info}). 

\begin{figure}
\epsscale{0.7}
\plotone{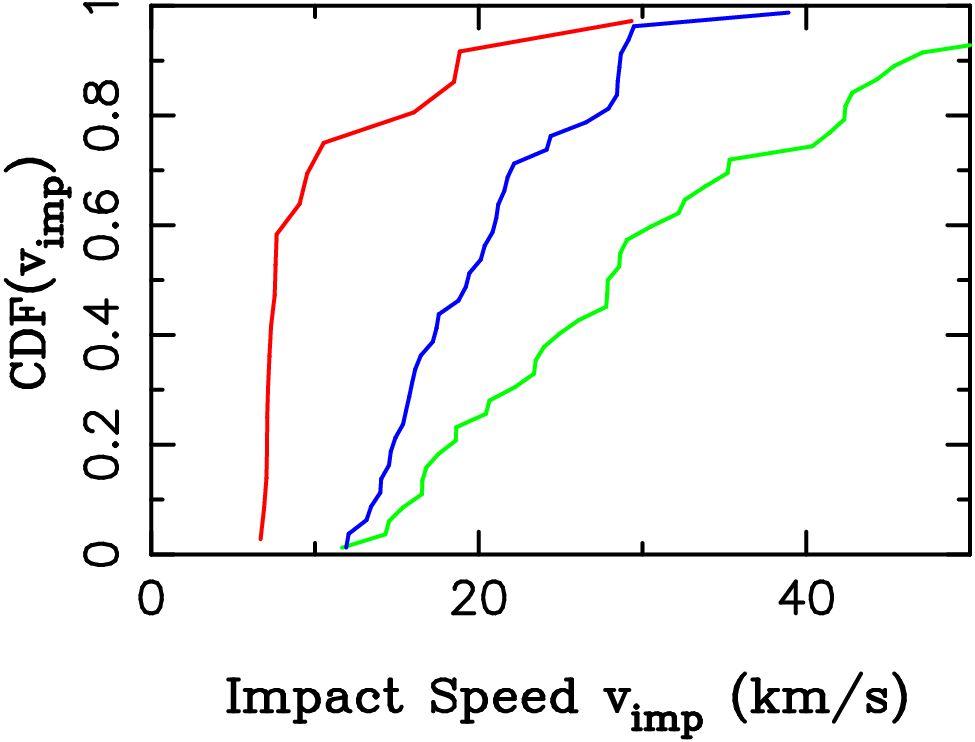}
\caption{The impact speed distributions of $D>5$ km asteroids on the terrestrial planets: Venus (green), Earth 
(blue), and Mars (red). The mean impact speeds are listed in Table~6. The median impact speeds are 27.9, 19.3 
and 7.6 km s$^{-1}$ for Venus, Earth and Mars, respectively.}
\label{vimp}
\end{figure}

\begin{figure}
\epsscale{0.9}
\plotone{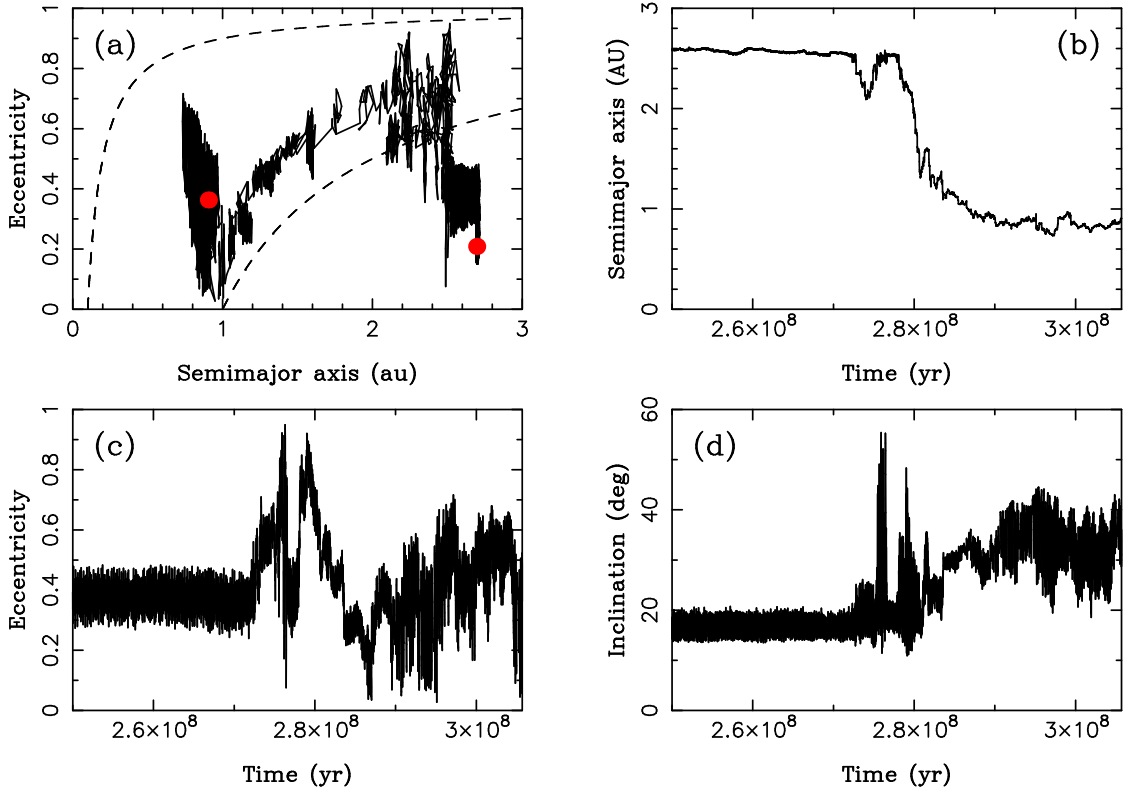}
\caption{The orbital history of asteroid 209192 ($D \simeq 5.7$ km, $p_{\rm V} \simeq 0.04$, assumed ${\rm d}a/{\rm d}t>0$) 
that started just below the 8:3 resonance and ended impacting on the Earth. The red dots in panel (a) show the 
initial and final orbits. We also plot $q=0.1$ au and $q=1$ au for a reference (dashed lines in (a)). This specific
MBA clone started sunward of the 8:3 resonance with Jupiter and drifted into the resonance with ${\rm d}a/{\rm d}t>0$.
It subsequently reached a Mars-crossing orbit and evolved to the 3:1 resonance with Jupiter by distant encounters 
with Mars. The 3:1 resonance transferred the asteroid to NEA space, where it impacted on the Earth at $t=306$ Myr.}
\label{ex1}
\end{figure}

The mean impact speeds are listed in Table 6 and the distribution of impact speeds is shown in 
Figure \ref{vimp}. We find that the mean impact speed of $D>5$ km NEAs on the Earth is nearly twice 
as large as the one on Mars (20.3 vs. 10.6 km s$^{-1}$). For Mars, about 60\% of impacts happen with 
the relatively low impact speeds $v_{\rm imp}<8$ km s$^{-1}$. Venus shows much larger impact speeds with 
$\sim 25$\% of impactors having $v_{\rm imp}>40$ km s$^{-1}$ (Fig. \ref{vimp}). Moreover, 3 out of 52 
Venus impactors (6\%) evolved to a retrograde orbit before the impact, and hit Venus at speeds 
exceeding 60 km s$^{-1}$. No such very-high-speed impact was recorded on the Earth but that is 
probably just an issue of small number statistics. The impact speeds of bright and dark asteroids 
are similar (e.g., terrestrial impacts with mean $v_{\rm imp}=20.5$ km s$^{-1}$ for bright and mean 
$v_{\rm imp}=20.0$ km s$^{-1}$ for dark). Dark asteroids often evolve to small orbital radii before 
they impact (see Fig. \ref{ex1} for an example). 

\section{Discussion}

\subsection{Comparison with Many-Source Models}

Compared to B02 and G18, here we find a larger contribution of the middle/outer belt to NEAs/impactors. 
For example, G18 estimated that the contribution of outer MBAs to NEAs is practically negligible 
($\simeq3.5$\% for the 2:1 resonance complex). They found that $\simeq 80$\% of impactors on the 
terrestrial worlds are produced from the $\nu_6$ resonance, and over 10\% from the 3:1 resonance, 
Hungarias and Phoaceas, leaving $<10$\% for the middle/outer belt. Based on this, G18 suggested that 
the majority of primitive NEOs/impactors come from the $\nu_6$ resonance. Here we find, instead, that 
the middle/outer belt supplies nearly 50\% of NEAs, $\simeq$70\% of dark NEAs, and $\simeq35$--40\% 
of large impactors.

These differences may be a consequence of the size-dependent delivery process. On one hand, small MBAs 
can drift over a considerable radial distance by the Yarkovsky effect and reach NEA space from the 
powerful $\nu_6$ resonance at the inner edge of the asteroid belt (e.g., Granvik et al., 2017). 
The $\nu_6$ resonance is known to produce highly evolved NEA orbits and impact probabilities on Earth 
in excess of 1\% (Gladman et al. 1997). On the other hand, large MBAs often reach NEA orbits via slow 
orbital evolution in weak resonances (Migliorini et al., 1998; Morbidelli and Nesvorn\'y, 1999; Farinella 
and Vokrouhlick\'y, 1999). Figure \ref{ex1} shows an example for the 8:3 resonance. Whereas each of these 
resonances contributes only a little, their total contribution to the population of large NEAs is 
significant. This can explain the more even contribution of different radial zones of the main belt 
to {\it large} NEAs.

Whether the transition from the $\nu_6$-controlled statistics for small asteroids to a more even 
contribution for large asteroids really happens for $1\lesssim D \lesssim 5$ km has yet to be established. 
Our results strictly apply for $D>5$ km. The single-source model could be extended to $D<5$ km, but 
this would require to develop a more realistic model for the YORP/Yarkovsky effects and collisions, 
and is left for future work. The many-source model of G18 was developed for $H>17$ ($D<1.7$ km for 
$p_{\rm V}=0.1$), where the model results can be calibrated on a large number of NEO detections by 
the Catalina Sky Survey (Christensen et al., 2012; Jedicke et al., 2016). The many-source model could 
be extended to $H<17$, but this is difficult to do with confidence because there are far fewer large 
NEO detections, and the calibration method would be affected by statistical uncertainties. 

Alternatively, the tension between the single- and many-source model results could be related to the 
time variability of the NEO population/impactor flux. The many-source models are calibrated on the 
{\it currently observed} population of NEOs, whereas our single-source model deals with the long-term 
average. There are all sorts of interesting issues that may arise from time variability. For example, 
it has been suggested that the cratering rate in the inner Solar System increased by a factor of 
$\sim$2-4 about 300-400 Myr ago (e.g., Culler et al., 2000; Mazrouei et al., 2019). If that is the 
case, we may be living in an epoch when the NEO population is $\sim$3 times larger than the 
long-term average (Fig. \ref{neapop}(a)). Catastrophic breakups of large parent MBAs near resonances 
can be responsible for changes of the NEA population/impactor flux. For example, the formation of the 
Flora family at the inner edge of the asteroid belt -- near the $\nu_6$ resonance --  was probably 
responsible for a factor of $\sim 2$ increase in the number of impacts 1-1.5 Gyr ago (Vokrouhlick\'y 
et al., 2017).  

Finally, some of the differences between the single- and many-source models may be a consequence of 
the adopted methodologies. The many-source models are agnostic to the radial distribution of MBAs (B02; 
Greenstreet et al., 2012; G18; Morbidelli et al., 2020). They do not take into account the availability 
of MBAs in (and near) different sources. This may potentially create a conflict between what is needed 
and what is available. For example, the many-source model can give a very strong weight to the $\nu_6$ 
resonance even if there are not enough asteroids near the $\nu_6$ resonance to justify it. One also 
needs to factor in that the outer belt has $\sim$10 times more MBAs than the inner belt (Sect. 5.1). 
The single-source model takes this into account but may have deficiencies elsewhere. For example, the 
simple physical model of radiation effects that we adopt here may fail to realistically emulate how 
MBAs reach resonances. Additional work is needed to improve our model and validate the results reported 
here. 

\subsection{Albedo and Taxonomy of Large/Small NEOs}

Observations of NEOs/MBAs in near-infrared reveal a bimodal distribution of visible albedos (e.g., Mainzer 
et al., 2011a, 2012; Masiero et al., 2012, 2014; Wright et al., 2016). The two albedo groups, roughly $p_{\rm V} 
\leq 0.1$ and $p_{\rm V} > 0.1$, which is the definition of dark and bright bodies that we adopt throughout 
this paper, are neatly correlated with the taxonomic classification of asteroids (Mainzer et al., 2011b; 
Pravec et al. 2012). The C-complex asteroids, representing primitive, carbonaceous-rich bodies thought to 
be implanted in the asteroid belt from the giant planet region (e.g., Walsh et al., 2011; Levison et al., 
2009), are dark. The S-complex asteroids, representing an alphabet soup of different asteroid classes 
related to ordinary chondrites, HEDs, and other types of stony meteorites, are typically bright. The mean 
albedos of C- and S-complex asteroids are $p_{\rm v}=0.057$ and 0.197, respectively (Pravec et al., 2012).

The albedo-based classification of different asteroid types is only approximate -- e.g., some S types can be 
dark if churned by impacts (Britt and Pieters, 1991) -- but it is adopted here as a rough guide. In 
Table 1, we show that 7 $D>5$ km NEAs that have been taxonomically classified as S (Binzel et al., 2019) 
all have $p_{\rm V}>0.1$, and 4 $D>5$ km NEAs that have been classified as C (3 cases) or D (1 case) all have 
$p_{\rm V} \leq 0.1$ (dark D-type NEOs on cometary orbits with $Q>4.5$ au are not considered here). The 
C-type asteroids, which are the dominant type among dark NEAs/MBAs (D and P asteroids are more common 
among Hildas and Jupiter Trojans; Emery et al., 2015), are thought to be the parent bodies of carbonaceous 
chondrite (CC) metorites (e.g., Clark et al. 2011). Here we therefore adopt a schematic view that: ``dark'' 
equals to ``primitive C type'' equals to ``carbonaceous chondrite''. The results reported here can 
therefore be used, with some caution, to understand the significance of CC material for NEAs and 
terrestrial impacts (Sect. 6.3).

We found that dark bodies represent $\sim 50$\% of $D>5$ km NEAs (and $\simeq 45$\% of $D>5$ km terrestrial 
impactors). Such a large share may be surprising because many previous works suggested that the 
primitive/dark NEOs should be less common. For example,  Mainzer et al. (2012) reported that NEAs 
with $p_{\rm V}<0.1$ represent $\simeq 39$\% of NEOs observed by WISE, and Binzel et al. (2019) reported 
that C-type NEAs represent only $\simeq 20$\% of taxonomically classified bodies. The statistic inferred 
from infrared observations is insensitive to the visible albedo (Mainzer et al. 2015), but it 
still contains an orbital bias, because dark NEAs typically have larger orbits than bright NEAs and are
therefore fainter and harder to detect in any wavelength (Fig. \ref{resid}). Complex selection, albedo 
and orbital biases affect the taxonomic observations as well. This means that the share of dark C-type
NEAs should be larger than reported in Mainzer et al. (2012) and Binzel et al. (2019). 

Morbidelli et al. (2020) investigated this issue in the many-source model of G18 and found that dark 
bodies ($p_{\rm V} \leq 0.1$) represent $\simeq57$\% of the total population of small NEOs (in the 
unbiased, size-limited sample). This is fully consistent with the results obtained here (Table 1 and 
Fig. \ref{neapop}(b)). The fact that small (Morbidelli et al. 2020) and large (this work) NEOs have 
roughly the same share of primitive dark bodies is intriguing. In Sect. 6.1 we argued that the 
single- and many-source models are giving different results because the statistics obtained in 
these models may depend on asteroid size. If the statistics were strongly size dependent, however, 
we would expect that dark bodies should represent a smaller share of small NEAs (than found here 
for large NEAs). But this is not the case.\footnote{Note that the contribution of different sources 
was set in G18 with no regard to the albedo distribution, and then used in Morbidelli et al. (2020) 
to determine the dark/bright split in each source (without re-fitting each source's {\it total} 
contribution). Simultaneously calibrating the many-source model on the Catalina and WISE observations 
of NEOs, however, would probably run into problems with degeneracies between different sources.}

\subsection{Implications for K/Pg-Scale Impacts}

The Cretaceous-Paleogene (K/Pg) boundary $\simeq 66$ million years ago corresponds to one of the three 
largest mass extinction events in the past 500 million years (Alroy 2008). It famously ended the age 
of dinosaurs. The K/Pg mass extinction is thought to have been triggered by a large asteroid impact 
(Alvarez et al., 1980) just off the coast of the Yucatan peninsula, forming a 180-km-wide Chicxulub 
crater (Schulte et al., 2010). The impactor that produced the Chicxulub crater is estimated to be at 
least 10 km in size hitting the Earth surface at a steep angle 45--60$^\circ$ to horizontal (Collins 
et al., 2020). Chromium found in sediment samples taken from different K/Pg boundary sites (Shukolyukov 
and Lugamir, 1998; Trinquier et al., 2006; Goderis et al., 2013), as well as a meteorite found in K/Pg 
boundary sediments from the North Pacific Ocean (Kyte et al., 1998), suggest the impactor was a 
CM-type carbonaceous chondrite (CC). This classification rules out the possibility that the K/Pg 
impactor came from an S-complex asteroid. 

The CC composition of the K/Pg impactor is surprising because the dominant extraterrestrial material 
hitting Earth appears to be ordinary chondrites (OCs; e.g., Maier et al., 2006; Koeberl et al., 2007, 
Tagle et al., 2009). For example, from about a dozen terrestrial craters with diameters $D_{\rm cra}=10$--100 
km and known (or suspected) impactor composition (Schmieder and Kring, 2020) only the Zhamanshin crater 
is believed to have been produced by a carbonaceous chondrite (Magna et al., 2017). In contrast, Chicxulub 
is one of only four recognized terrestrial craters with $D_{\rm cra}>100$ km, the other three being 
Popigai, Sudbury and Vredefort. The 36.6-Myr old, $D_{\rm cra} \simeq 100$ km Popigai crater in central 
Russia was produced by an OC impactor (Tagle and Claeys, 2005). The impactor types for the much larger
and older Sudbury and Vredefort craters are uncertain, but if the Zaonega spherule layers are linked  
to the Vredefort impact (Mougel et al., 2017), they would indicate a CC composition of the impactor.
In summary, $\sim25$--75\% of the largest craters on the Earth were produced by CC impactors.  

The inferred composition of the K/Pg impactor is an important clue to its origin. It has been 
suggested, for example, that the K/Pg impactor may have been a fragment of a large, inner-main-belt asteroid 
that catastrophically disrupted 100-200 Myr ago, and left behind the Baptistina family (Bottke et al., 2007; 
Masiero et al., 2012). Subsequent infrared and spectroscopic observations revealed, however, that the 
Baptistina family probably does not have the right composition ($p_{\rm V} \simeq 0.18$ albedo from WISE 
and S-type spectrum; Reddy et al., 2011). The flux of cometary impactors is negligible compared
to asteroids (e.g., dormant JFCs contribute by $<$1\%; G18). Siraj and Loeb (2021) proposed that the 
impactor was a piece of a large long-period comet that tidally disrupted near its perihelion. This 
exotic idea has a number of problems, including the very low likelihood of terrestrial impacts from 
the long-period comets, the low efficiency of tidal disruption to make $D \gtrsim 10$ km fragments, 
and the general assumption that the geochemical signature of a cometary impactor would be consistent 
with the CM composition inferred for the K/Pg impactor.

Here we argue that the problem at hand may have a simple solution, because nearly a half of {\it large} 
terrestrial impactors are dark primitive asteroids with composition that is consistent with carbonaceous 
chondrites (via the relationship of low asteroid albedo to the C-type taxonomy and CC composition; 
e.g., Mainzer et al., 2011a; Pravec et al., 2012; Binzel et al., 2019).  By modeling the 
orbital evolution of large MBAs, we show that large asteroids escape from the main belt via weak 
resonances (see Fig. \ref{ex1} for an example). The large NEAs more evenly sample the radial extension 
of the main belt (Sect. 5.3), including the middle/outer belt where dark, C-type asteroids are quite 
common (Sect. 5.1). That is why the large terrestrial impactors are often (in $\simeq 45$\% of cases; 
Table~5) dark primitive bodies with a (plausible) CC composition. Small, $D \lesssim 1$ km terrestrial 
impactors have predominantly OC composition because their delivery from the main belt is controled 
by the powerful $\nu_6$ resonance at the inner edge of the asteroid belt (where the OC material 
is common; Sect 5.1).

Our model with random Yarkovsky drifts indicates $\sim 2$ $D>10$ km asteroid impacts on the Earth in 
1 Gyr. This is about a factor of 2 higher than the impact flux inferred in Nesvorn\'y and Roig (2018), 
possibly because the \"Opik code used in that work had underestimated the number of impacts. The NEA 
population and impact rates tend to be up to a factor of $\sim 2$ higher with optimized Yarkovsky drifts 
(Sect. 5.2 and 5.5). We therefore find that $\sim$2-4 impacts of $D>10$ km asteroids should happen 
on the Earth in 1 Gyr and that the average spacing between impacts should be $\sim0.25$--0.5 Gyr 
This is a useful input for understanding the frequency of impact-related mass extinctions on the Earth.
About half of these large impactors are carbonaceous chondrites (see above). Having at least one mass 
extinction from a $D>10$ km impact happening in the last 100 Myr, for example, is a roughly a 20-40\% 
probability event, or a 10-20\% probability event if the CC composition of the Chicxulub impactor 
is factored in. 

\section{Conclusions}

We conducted dynamical simulations of known $D>5$-km main-belt asteroids as they evolve onto NEA orbits by 
radiation forces and resonances. The results were used to develop a single-source model for large NEAs 
and terrestrial impactors. The main advantage of the single-source model is that it takes into 
account the availability of MBAs near source resonances. Our main findings are:
\begin{enumerate}

\item The NEA model with random Yarkovsky drifts is incompatible with the number of observed $D>5$
km NEAs ($n_5=17$; Table 1). This could mean that the current population of large NEAs is a factor of 
$\sim$2-3 larger than the long-term average (Mazrouei et al. 2019). In the model with Yarkovsky 
optimized drifts there is a $\simeq$11\% probability to have $\geq 17$ $D>5$ km NEAs at any single
epoch. 

\item The long-term average of the ratio of dark ($p_{\rm V} \leq 0.1$) and bright ($p_{\rm V} > 0.1$) 
$D>5$ km NEAs peaks near 0.8 and the mean value is 1.3. There is a good agreement with 
current observations which show 8 dark and 9 bright $D>5$ km NEAs (i.e., $\sim0.9$ ratio). 

\item Nearly 50\% of diameter $D>5$ km NEAs start in the middle/outer belt ($a>2.5$ au) and reach 
NEA orbits via weak orbital resonances (e.g., the 8:3 resonance with Jupiter at 2.7 au). The middle 
belt is the main contributor of dark NEAs (50\%) and the inner belt is the main contributor of bright 
NEAs (76\%). 

\item The contribution of asteroid families to $D>5$ km NEAs is relatively minor. We find that only 
20\% of $D>5$ km NEAs in a steady state are expected to originate in known families. The remaining 80\% 
are background MBAs. The family contributions to dark and bright $D>5$ km NEAs are the same. 

\item Our model fails to match -- but only at 2--3$\sigma$ -- the semimajor axis and perihelion distance 
distributions of observed $D>5$ km NEAs. For $p_{\rm V} \leq 0.1$, the model suggests that $\sim$40\% 
of large NEAs should have $a<2$ au or $q<0.7$ au, but all 8 known $D>5$ km NEAs with $p_{\rm V} \leq 0.1$ 
have $a>2$ au and $q>0.7$ au. We investigated models where NEAs with $q<q^*$ are removed (e.g., disrupted 
by thermal stresses) and found $0.1 \lesssim q^* \lesssim 0.3$ au would provide a better match to observations.

\item The time-averaged impact flux of $D>5$ km asteroids on the Earth is $25 \pm 7$ Gyr$^{-1}$, where the 
uncertainty expresses the dependence of the model results on the distribution of MBA spin vectors (that  
influence the Yarkovsky drift and influx of MBAs on NEA orbits). Venus receives about the same number 
of impacts as the Earth, and Mars receives $\simeq2.7$ fewer impacts than the Earth. 

\item MBAs starting with $a>2.5$ au produce 35--40\% of $D>5$ km asteroid impacts on the terrestrial 
planets. For comparison, G18 reported that $\simeq80$\% of the terrestrial impactors with absolute magnitudes 
$17<H<22$ ($0.17<D<1.7$ km for $p_{\rm V}=0.1$) come from the $\nu_6$ resonance at the inner edge of the 
main belt. This suggest that the contribution of different parts of the main belt to terrestrial impacts 
is size dependent.

\item We find that 47\% of large terrestrial impactors are dark ($p_{\rm V}\leq0.1$) and 53\% are bright 
($p_{\rm V} > 0.1$), suggesting a roughly equal split. About 60\% of dark impactors come from the 
middle/outer belt ($a>2.5$ au). In contrast, according to G18, most small/dark impactors on the Earth 
come from the inner belt.

\item Nearly 80\% of bright $D>5$ km impactors start in the inner belt ($a<2.5$ au). Of these, 18\% were 
previous members of the Flora family. In total, the asteroid families contribute by 23\% to impacts 
of $D>5$ km asteroids on the terrestrial worlds. 

\item The number of impacts on the Earth impacts drops to 67\% for $q^*=0.1$ au and 51\% for 
$q^*=0.3$ au (both percentages quoted with respect to the original impact flux without the NEA removal). 
The removal effect on impacts is stronger for Venus (48\% for $q^*=0.1$ au and 31\% for $q^*=0.3$ au), 
and negligible for Mars. 

\item The mean (median) impact velocities of large NEAs are 30.5 (27.9), 20.3 (19.3) and 10.6 (7.6) km 
s$^{-1}$ for Venus, Earth and Mars, respectively.     
 
\end{enumerate}

The findings reported here have important implications for our understanding of the occurrence rate 
and nature of the K/Pg-scale impacts on the terrestrial worlds. Impacts of $D>10$ km asteroids on the Earth 
do not happen often (average spacing $\sim 250$--500 Myr), implying that having one K/Pg-scale event in the 
past 100 Myr was mildly special ($\sim10$--40\% probability; see discussion in Sect. 6.3). About a half of 
large terrestrial impactors are expected to be dark carbonaceous asteroids. 
  
\acknowledgments
This work was supported by NASA's SSW program. The simulations were performed on NASA's Pleiades Supercomputer.  
We greatly appreciate the support of the NASA Advanced Supercomputing Division. We thank A. Morbidelli for 
helpful discussions. We thank S. Raymond and the anonymous reviewer for comments on the submitted manuscript.

\begin{table}
\centering
{
\begin{tabular}{lrrrrrrr}
\hline \hline
number & name/desig. & $a$       & $e$ & $i$       & $D$  & $p_{\rm V}$ & tax. \\  
       &             & (au)      &     & ($^\circ$) & (km) &           &       \\
\hline
433    &  Eros       & 1.458 & 0.223 & 10.8 & 16.8 & 0.25 & S \\
1036   &  Ganymed    & 2.663 & 0.534 & 26.7 & 36.5 & 0.23 & S \\ 
1580   &  Betulia    & 2.197 & 0.488 & 52.1 & 6.5 & 0.07 & C \\
1627   &  Ivar       & 1.863 & 0.397 &  8.5 & 8.1 & 0.13 & S \\ 
1866   &  Sisyphus   & 1.893 & 0.539 & 41.2 & 6.6 & 0.26 & S \\ 
1917   &  Cuyo       & 2.149 & 0.506 & 24.0 & 5.0 & 0.20 & - \\
2212   &  Hephaistos & 2.160 & 0.838 & 11.6 & 5.5 & 0.16 & S \\
4954   &  Eric       & 2.002 & 0.449 & 17.4 & 8.9 & 0.19 & S \\
5731   &  Zeus       & 2.263 & 0.654 & 11.4 & 6.9 & 0.02 & - \\
7092   &  Cadmus     & 2.535 & 0.698 & 17.8 & 6.3 & 0.05 & - \\
20826  &  2000 UV13  & 2.425 & 0.633 & 31.9 & 5.9 & 0.12 & - \\
25916  &  2001 CP44  & 2.561 & 0.498 & 15.7 & 5.8 & 0.22 & S \\
26760  &  2001 KP41  & 2.853 & 0.557 & 11.0 & 5.4 & 0.04 & C \\
52762  &  1998 OR2   & 2.418 & 0.652 & 33.9 & 6.7 & 0.05 & D \\
88263  &  2001 KQ1   & 2.097 & 0.432 & 38.8 & 5.1 & 0.04 & - \\  
89830  &  2002 CE    & 2.077 & 0.507 & 43.7 & 5.1 & 0.08 & - \\
162566 &  2000 RJ34  & 2.636 & 0.574 & 13.9 & 6.0 & 0.03 & C \\
\hline \hline
\end{tabular}
}
\caption{A list of $D>5$ km NEAs from WISE (Mainzer et al., 2019). 
The taxonomic type in the last column was taken from Binzel et al. (2019). In the cases where more than 
one albedo/diameter measurements were available from WISE, we report the arithmetic mean of these measurements. 
Given the reported uncertainty of WISE-inferred size estimates ($\sim20$\% 1$\sigma$ uncertainty; Nugent et 
al. 2016), some of these NEAs may actually have $D<5$ km. 
This is especially relevant for 1917 (bright), 88263 (dark) and 89830 (dark). We include these cases here for 
completeness. 3200 Phaethon, the target of future JAXA's DESTINY mission, has $D \simeq 4.2$ km from AKARI 
(Usui et al., 2011), $D \simeq 4.6$ km from WISE (Masiero et al., 2019), $D \simeq 5.1$ km from 
Spitzer (Hanu\v{s} et al., 2016), and $D \simeq 5.8$ km from occultation observations (Dunham et al. 2020). 
Phaeton is not included here but we briefly discuss it in Sect. 5.4. The data 
for (433) Eros were obtained from the NEAR-Shoemaker mission (Yeomans et al., 2000). The known NEOs with 
$Q=a(1+e)>4.5$ au are not listed here because they are likely dormant JFCs: 3552 Don Quixote, 
5370 Taranis, 2014 JH57 and 2014 MQ18 -- all are dark, 3552 is a D type.}
\end{table} 


\begin{table}
\centering
{
\begin{tabular}{lrr}
\hline \hline
domain & range        &  lifetime    \\
       &              & (Myr)        \\ 
\hline       
all    & $1.7<a<4.5$ au     & 1.2    \\
inner  & $a<2.5$ au         & 6.5    \\
Flora/$\nu_6$ & $a<2.3$ au  & 8.2    \\
3:1    & $2.45<a<2.55$ au   & 2.9    \\
middle & $2.5<a<2.82$ au    & 2.2    \\
8:3    & $2.65<a<2.75$ au   & 2.3    \\   
5:2    & $2.77<a<2.87$ au   & 0.66   \\      
outer  & $a>2.82$ au        & 0.54   \\
\hline
dark   & $p_{\rm V}\leq0.1$  & 0.93   \\
bright & $p_{\rm V}>0.1$     & 2.1    \\
\hline \hline
\end{tabular}
}
\caption{The dynamical lifetime of NEAs. The mean dynamical lifetime was determined from our model 
(Sect. 4.4) for different source regions and different NEA types (two bottom rows). The semimajor axis
and albedo range in the second column defines the source domains.}
\end{table}

\begin{table}
\centering
{
\begin{tabular}{lrrr}
\hline \hline
region      & range        & optimized                &  random    \\  
\hline    
Hungarias & $a<2$ au        &  4.5\%                  & 3.8\%      \\ 
inner  & $a<2.5$ au         &   {\bf 52\%}            &  {\bf 50\%}  \\
Phoaceas & $a<2.5$ au, $i>17^\circ$ & 9.1\%            &  8.2\%     \\ 
inner, no 3:1 & $a<2.45$ au &   46\%                  &  45\%      \\
3:1    & $2.45<a<2.55$ au   &   15\%                  &  12\%      \\
middle & $2.5<a<2.82$ au    &   {\bf 35\%}            &  {\bf 34\%}      \\
5:2    & $2.77<a<2.87$ au   &    1.3\%                &  2.1\%       \\      
outer  & $a>2.82$ au        &   {\bf 13\%}            &  {\bf 15\%}  \\  
Euphrosyne & $3<a<3.3$ au, $24^\circ<i<32^\circ$ & {\it 3.2\%}  &  {\it 6.8\%}   \\ 
\hline \hline
\end{tabular}
}
\caption{The contribution of different source regions to $D>5$ km NEAs. The values listed in 
columns 3 and 4 are the percentages obtained from the models with optimized and random drifts, 
respectively. The model with optimized drifts is more consistent with the number of observed 
$D>5$ km NEAs (Fig. \ref{neapop}(a)).}
\end{table} 

\begin{table}
\centering
{
\begin{tabular}{lrrr}
\hline \hline
region      & range      & dark                &  bright          \\  
            & (au)           & $p_{\rm V} \leq 0.1$  & $p_{\rm V} > 0.1$ \\
\hline
inner  & $a<2.5$       &   29\% (26\%)      &  76\% (75\%)     \\
middle & $2.5<a<2.82$  &   50\% (47\%)      &  21\% (22\%)     \\
outer  & $a>2.82$      &   21\% (28\%)      &  3\%  (3\%)      \\ 
\hline \hline
\end{tabular}
}
\caption{The contribution of different source regions to dark and bright $D>5$ km NEAs. The two values
listed in each case were obtained for the optimized and random drift models, with the latter estimate 
given in parentheses.}
\end{table} 

\begin{table}
\centering
{
\begin{tabular}{lrrrrr}
\hline \hline
domain & range             & Venus       &  Earth        & Mars  &    all     \\  
\hline
inner  & $a<2.5$ au        &  56\%       &  65\%        & 61\%      &    60\%    \\
middle & $2.5<a<2.82$ au   &  31\%       &  18\%        & 28\%      &    25\%    \\
outer  & $a>2.82$ au       &  13\%       &  17\%        & 11\%      &    15\%    \\ 
\hline
dark   & $p_{\rm V}\leq0.1$ &   44\%      &   45\%        & 61\%     &    47\%    \\
bright & $p_{\rm V}>0.1$    &   56\%      &   55\%        & 39\%     &    53\%     \\
\hline \hline
\end{tabular}
}
\caption{The contribution of different source/albedo domains to impacts. In total, the $N$-body integrator 
recorded 52 Venus impacts, 49 Earth impacts, and 18 Mars impacts in 1 Gyr.}
\end{table} 

\begin{table}
\centering
{
\begin{tabular}{lrr}
\hline \hline
       & $v_\infty$ & $v_{\rm imp}$ \\  
       & (km s$^{-1}$)      & (km s$^{-1}$)    \\
\hline
Venus  &  28.2       & 30.5 \\
Earth  &  16.4       & 20.3 \\
Mars   &   9.0       & 10.6 \\ 
\hline \hline
\end{tabular}
}
\caption{The mean impact speed of $D>5$ km asteroids on the terrestrial planets: $v_\infty$ is the velocity
in `infinity' (i.e., before focusing), and $v_{\rm imp}$ is the actual impact speed. For reference, the escape 
speeds are 10.3 (Venus), 11.2 (Earth), and 5.0 km s$^{-1}$ (Mars).}
\end{table} 

\clearpage

\clearpage

\clearpage

\clearpage

\clearpage

\clearpage

\clearpage

\clearpage

\end{document}